\documentclass[a4paper,11pt]{article}
\usepackage{pos}

\usepackage{bbm}

\usepackage{amsmath}
\usepackage[all]{xy}

\usepackage{tikz}
\usetikzlibrary{cd}

\DeclareMathAlphabet{\mathpzc}{OT1}{pzc}{m}{it} 

\definecolor{darkblue}{rgb}{0.05,0.25,0.65}
\definecolor{darkgreen}{RGB}{20,140,10}
\definecolor{lightgray}{rgb}{0.9,0.9,0.9}
\definecolor{darkorange}{RGB}{200,100,5}
\definecolor{darkyellow}{rgb}{.91,.91,0}
\definecolor{lightolive}{RGB}{189,183,107}

\definecolor{greenii}{RGB}{20,140,10}
\definecolor{orangeii}{RGB}{200,100,5}

\usepackage{multirow}

\usepackage{enumerate} 

\theoremstyle{italics}
\newtheorem{theorem}{Theorem}[section]

\theoremstyle{definition}
\newtheorem{definition}[theorem]{Definition}

\newtheorem{example}[theorem]{Example}

\newtheorem{remark}[theorem]{Remark}

\newcommand{\CartesianSpaces}{\mathrm{CartSp}}

\newcommand{\SmoothManifolds}{\mathrm{SmthMfd}}

\newcommand{\SuperManifolds}{\mathrm{SupMfd}}

\newcommand{\SmoothSets}{\mathrm{SmthSet}}

\newcommand{\SuperSmoothSets}{\mathrm{Sup}\SmoothSets}

\newcommand{\OddCartesianSpaces}{\mathrm{Odd}\CartesianSpaces}

\newcommand{\SuperSets}
{\mathrm{SupSets}}

\newcommand{\SuperCartesianSpaces}{\mathrm{SupCartSp}}

\newcommand{\CX}{\mathcal{X}}

\newcommand{\CG}{\mathcal{G}}

\newcommand{\dd}{\mathrm{d}}

\newcommand{\CY}{\mathcal{Y}}

\newcommand{\CL}
{\mathcal{L}}

\newcommand{\CF}
{\mathcal{F}}

\newcommand{\CE}
{\mathcal{E}}

\newcommand{\id}{\mathrm{id}}

\newcommand{\FR}{\mathbbm{R}}

\newcommand{\odd}
{\mathrm{odd}}

\newcommand{\CO}
{\mathcal{O}}

\newcommand{\DD}
{\mathbbm{D}}

\newcommand{\epsi}
{\mathcal{\epsilon}}

\title{Sheaf Topos Theory: \\ A powerful setting for Lagrangian Field Theory}

\author*[a]{Grigorios Giotopoulos}

\affiliation[a]{Mathematics, Division of Science; and \\ 
Center for Quantum and Topological Systems,
NYUAD Research Institute,\\
New York University Abu Dhabi, UAE.}

\emailAdd{grigorios.giotopoulos@nyu.edu}

\abstract{We provide an introductory exposition to the sheaf topos theoretic description of classical field theory motivated by the rigorous description of both $\bf{(i)}$ the variational calculus of (infinite dimensional) field-theoretic spaces, and $\bf(ii)$ the non-triviality of classical fermionic field spaces. These considerations naturally lead to the definition of the sheaf topos of super smooth sets. We close by indicating natural generalizations necessary to include to the description of infinitesimal structure of field spaces and further the non-perturbative description of (higher) gauge fields.}

\FullConference{Proceedings of the Corfu Summer Institute 2024 "School and Workshops on Elementary Particle Physics and Gravity" (CORFU2024)\
12 - 26 May, and 25 August - 27 September, 2024\\
Corfu, Greece\\}

\tableofcontents

\begin{document}
\maketitle

\section{Introduction}\label{Introduction}
Most modern considerations in theoretical physics assume that the physical world is : \\ (1) field-theoretic, (2) smooth, (3) containing fermions, (4) local,  (5) gauged, and
last but not least \\ (6) non-perturbative. In more detail, this is means that nature is to be described fundamentally by \textit{bosonic} and \textit{fermionic fields}, whose dynamics are prescribed by \textit{local} Langrangians and (classically) by their induced (on-shell) Euler--Lagrange partial differential equations. The totality of (off-shell) field configurations is implicitly assumed to be a (necessarily infinite-dimensional) \textit{``smooth space''}, in such a manner so that one may perform the standard operations of variational calculus \textit{as if} it were a finite-dimensional manifold. When fermions are included, this ``smooth space'' should be such that the \textit{anticommuting nature} of fermionic fields is manifest \textit{as if} they were odd coordinates on a (finite-dimensional) super-manifold\footnote{Although obvious to experts, we stress here that this is already necessary at the \textit{kinematical} level, that is, even if supersymmetry is absent in the dynamics under study.}.  Moreover, when gauge fields are taken into account the (smooth) field space should manifestly encode (finite/non-perturbative) \textit{gauge transformations} as (internal) symmetries ($\equiv \,$ \textit{isomorphisms}) between gauge equivalent field configurations\footnote{Accordingly, considerations of higher gauge fields dictate that the corresponding field space should also manifestly encode \textit{higher gauge transformations} as higher (internal) symmetries ($\equiv$ \textit{higher morphisms}) between lower order gauge transformations and so on.}. Finally, this notion of field space should also be \textit{non-perturbative}, i.e.,  naturally encode all \textit{non-trivial topological sectors} of field configurations\footnote{Since these exist in nature, i.e., as monopoles, instantons, solitons, skyrmions etc.}, by necessity towards the hope of constructing a non-perturbative quantization / path integral scheme.

This exposition may be viewed as an introduction to the series recently initiated with \cite{GS25}, to be followed by \cite{GSb}\cite{GSc} and \cite{GSSd}, with the aim of rigorously describing all of these aspects under a common mathematical theory, in a manner that is approachable by both mathematicians and theoretical physicists\footnote{This series is in turn inspired by the ideas of Schreiber \cite{dcct} and the early live lecture notes \cite{Schreiber} with a
viewpoint towards perturbative quantum field theory.}. Along the way, we provide a partial survey of results that have appeared in \cite{GS25}, and further results that will appear in the following parts of the series.

Elaborate mathematical frameworks have  already been developed to describe \textit{separately} each of these aspects. For instance, an appropriate\footnote{Appropriate in that, at the very least, it satisfies the ``Exponential Law''; Namely, it should be such that there is a canonical identification of smooth maps $\{f: \Sigma \rightarrow C^\infty(M,N)\}$, from any manifold $\Sigma$, with smooth maps of manifolds $C^\infty(\Sigma\times M\, ,\, N)$.} notion of smoothness of bosonic field spaces, i.e., $\sigma-$model mapping spaces of the form 
\vspace{-1mm}
$$
C^\infty(M\,,\,N)
$$

\vspace{-1mm}
\noindent
where $M,N$ are two smooth manifolds, or more generally spaces of sections 
\vspace{-1mm}
$$
\Gamma_M(F)
$$

\vspace{-1mm}
\noindent
of a fiber ``field bundle'' $F\rightarrow M$, has been formalized via Fr\'{e}chet manifold theory \cite{KrieglMichor}\cite{Hamilton}). This involves elaborate functional analysis which is mostly useless for practicing theoretical physicists. Moreover, it actually applies \textit{only} when the base spacetime $M$ is compact\footnote{Generalizations of such smooth structures via (non-Fr\`{e}chet) infinite-dimensional charts exist for the case of non-compact $M$, but are burdened with further technicalities and unnatural choices \cite[Ch. IX]{KrieglMichor}.}, which is clearly undesirable from a physical perspective. On the other hand, the locality aspect of field theory has been pinpointed in the structure of the 
\textit{infinite jet bundle} 
$$
J^\infty_M F
$$ 
and its corresponding variational bi-complex $\Omega^{\bullet,\bullet}(J^\infty_M F)$ of differential forms \cite{Anderson89}\cite{Saunders89}. The standard (algebraic) manipulations of variational calculus for bosonic fields may be delegated to corresponding operations on the variational bi-complex, with the resulting expressions being ``pulled back'' to the field space via the jet prolongation map
$$
j^\infty \, : \, \Gamma_M(F) \longrightarrow \Gamma_M(J^\infty_M F) \, .
$$
Nevertheless, the infinite dimensional jet bundle $J^\infty_M F$ has traditionally  been treated only as a formal space, i.e., a formal limit of finite dimensional smooth manifolds (the finite jet bundles $J^k_M F$), simply by declaring what the set of smooth real-valued functions $C^\infty(J^\infty_M F)$ on it should be. That is, it has not been treated as a smooth space on the same footing as the field spaces which are of actual physical interest\footnote{It is true, however, that the infinite jet bundle may be treated a special kind of a Fr\'{e}chet manifold, even if $M$ is non-compact \cite{Takens79}\cite{KS17}\cite{GS25}.}. As we shall indicate in Sec. \ref{SmoothSetsSec}, all of these difficulties are circumvented if one instead works within the sheaf topos of $\textit{Smooth Sets}$, while also faithfully subsuming the Fr\'{e}chet descriptions, following the fully detailed account given in \cite{GS25}.

Analogously, field spaces that include fermions have been advocated to be treated as certain kinds of infinite-dimensional supermanifolds \cite{Schmitt97} \cite{Hanisch14}, but this approach is burdened by the same and \textit{more} technical difficulties as in the purely bosonic case. Of course, the main conceptual issue for fermions at the classical level is to make mathematical sense of the \textit{anticommuting symbol} $\psi$, appearing for instance in expressions such as the Dirac Lagrangian
$$
\CL_{\mathrm{Dirac}} \, = \, (\overline{\psi} \,\gamma^\mu  \partial_\mu \psi ) \cdot  \dd^4x 
$$
entering the total Standard Model Lagrangian, or even that of the dimensionally reduced toy example of a fermionic particle on the real (time) line
$$
\CL_{\mathrm{Fer. Part.}} \, = \, \psi
\, \partial_t \psi \cdot \mathrm{d} t \, , 
$$
and further any polynomial expression involving $\psi$ and its derivatives (such as ``observables'').  An alternative route to make sense of fermionic field spaces and such expressions, which is closer to the practice of the physics literature, has been via supplying arbitrary amount of ``\textit{auxilliary odd coordinates}'' in an appropriate ``functorial'' manner \cite{Molotkov84} (following \cite{DeWitt}\cite{Rogers07} in finite dimensions). The mathematical origin of this approach has been given a rigorous ``functor of points'' explanation in \cite{Sachse08} (see also \cite{Freed99} for motivation), in a similar vein to what we shall describe. We note, however, the fact  that this approach to the fermionic structure of field spaces should mathematically be on precisely the same level as their smooth structure was only recently amplified in \cite[\S 4.6]{dcct}\cite[\S 3.1]{Schr21-HPG} (to be fully developed in \cite{GSc}). Indeed, as we will briefly describe in Sec. \ref{SuperSmoothSetsSec}, working directly within the sheaf topos of \textit{Super Smooth Sets} bypasses all such issues in a straightforward manner.

Having made the step to describe the fermionic structure of field spaces, it turns out that making mathematical sense of the \textit{infinitesimals} as commonly used in physics (e.g. in  defining tangent vectors on manifolds, infinitesimal transformations of fields and perturbation theory via (nilpotent) formal parameters), is straightforward and follows analogously. We briefly indicate how this works and mention a few rigorous results obtained in this setting in Sec. \ref{GeneralizationSec}, with the full details to be brought forth in \cite{GSb}.

Regarding the rigorous description of internal symmetries, it has become by now clear that the natural framework for describing non-perturbatively (higher) gauge fields is that of (higher) \textit{groupoids} (see e.g. \cite[\S 2]{Schr21-HPG} for gauge fields as groupoids and \cite{Weinstein96} for a a general exposition to groupoids). For the purposes of field theory, these can similarly be augmented in the context of (higher) sheaf topos theory with the necessary smooth and super structure, i.e., as (higher) \textit{Super Smooth Groupoids}. We indicate the basics of how this works in an intuitive manner in Sec. \ref{GeneralizationSec}; we do so by avoiding the technical details involved in this short exposition  (for more details see \cite{dcct} \cite{Nesin}\cite{FSS23Char} \cite{Schreiber24}, to be fully expanded in \cite{GSSd}). 
\subsection*{Intuitive approach to sheaf topos theory}

The common mathematical framework that naturally incorporates all of the above aspects -- ``\textit{sheaf topos theory}'' -- has been implicitly known to a few experts in mathematical physics, but as far as we are aware it has not been fully and explicitly expanded upon in a manner that is accessible to the theoretical physicist's practice. Here we immediately give away the main conceptual shift in intuition which then allows one to leisurely follow the development of the theory. Namely, one should firstly revise what the core notion of a ``space'' should comprise of. Following the original idea of Grothendieck (see e.g. \cite{Grothendieck73}, then in the context of algebraic geometry), we shall advocate studying field theoretic spaces via what is commonly known in mathematics as the ``functor of points approach''. 

In simple terms, one should \textit{not} in general restrict a space to be a set of points supplied with extra structure (cf. topological spaces, smooth manifolds, super manifolds etc.). Instead, a geometrical space should be defined entirely operationally, by giving meaning to what it means to ``probe'' the would-be space with simpler test spaces whose nature we are already familiar with. In other words we shall know, or better \textit{define} such a generalized space $\mathcal{X}$ by \textit{consistently} answering the question:
\begin{quote}
 \textit{``What are the ways we can probe the would-be space $\mathcal{X}$ with a collection of simple probe spaces $C$?"}
 \end{quote} 

We will see that this change in perspective will naturally yield a rigorous definition, which will naturally incorporate the field spaces appearing in physics as such generalized spaces. Before that, however, we should note immediately that this manner of thinking is not far from that of experimental physics (e.g. probing materials with particles etc.). Moreover, it is also quite similar to the intuition arising in string theory, where a background spacetime should be ``whatever the particles/strings/branes detect'' as they transverse through it. Let us now show how careful consideration of the above proposition directly results in the definition of such a generalized notion of space as a ``\textit{(pre)-sheaf}'' on a chosen category\footnote{Recall, a category $C$ consists of a collection of \textit{objects} $C_0$ together with a collection of \textit{morphisms}  (i.e., directed arrows) $\mathrm{Hom}_C(\Sigma_1,\, \Sigma_2)$ for any two objects $\Sigma_1,\Sigma_2 \in C_0$, such that identity morphisms exists and one may (associatively) compose morphisms with common target and source. For an introduction to the basics of categories, functors and (co)limits see for instance \cite[\S 2]{Geroch85} which is aimed at mathematical physicists. Further details may be found in lecture notes such as \cite{Awodey06}\cite{Nesin} and also [\href{https://ncatlab.org/nlab/show/geometry+of+physics+--+categories+and+toposes}{\tt nlab : Geometry of physics-categories and toposes}].} of probe spaces.

\medskip 
\noindent {\bf Generalized spaces as Sheaves on a category of probes}
\begin{itemize}
\item
Firstly, to say we understand the nature of a collection of probe spaces means that we know how to map between them while ``preserving whatever internal structure they have''. In mathematical terms, we assume our probes $C$ form a category. For instance, this may be the category $\mathrm{Top}$ of topological spaces with morphisms being continuous maps, and more appropriate to our context the category of smooth manifolds $\SmoothManifolds$ with morphisms being smooth maps, or supermanifolds $\SuperManifolds$ with morphisms being maps of supermanifolds; or even their full subcategories consisting of only the corresponding model Cartesian spaces $\CartesianSpaces \hookrightarrow \SmoothManifolds$ and super-Cartesian spaces $\SuperCartesianSpaces \hookrightarrow \SuperManifolds$ (i.e., local charts). Although we will phrase the discussion below abstractly for an arbitrary category of probes (towards our end-goal of further generalizations indicated in Sec. \ref{GeneralizationSec}), the physics inclined reader may keep in mind the category $C$ of probes to be either of the familiar $\CartesianSpaces$ or $\SuperCartesianSpaces$. Indeed, these examples of probes suffice both for intuition and the field theoretic spaces we shall describe in Sec. \ref{SmoothSetsSec} and Sec. \ref{SuperSmoothSetsSec}.

\item Next, to say we know ``the ways that we can probe'' our would-be space $\mathcal{X}$ with a probe space $\Sigma \in C$, we are required at the bare minimum to prescribe a \textit{set}
\vspace{-1mm}
\begin{equation}\label{SetOfPlots}
\mathcal{X}(\Sigma) \quad \equiv \quad \mathrm{Plots} (\Sigma,\, \mathcal{X}) \quad \in \quad \mathrm{Set} \, ,
\end{equation}

\vspace{-1mm}
\noindent
for each $\Sigma \in C$. Of course, we think of these as mappings of $\Sigma$ into $\mathcal{X}$, hence identifying the \textit{set of $\Sigma$-shaped plots in $\mathcal{X}$}. Suggestively, we schematically denote elements of these sets
\vspace{-1mm}
$$
\phi_\Sigma \quad \in \quad  \CX(\Sigma) \, , 
$$
i.e., $\Sigma$-shaped plots, as arrows\footnote{For the moment these are not actual morphisms in some category, since as of yet the probe spaces and the would-be generalized spaces have not been identified to inhabit a common category, hence the quotation marks.} from $\Sigma$ to $\CX$
\begin{equation}
  \label{SchematicsOfProbes}
  \begin{tikzcd}[column sep=large]
     `` \, \phi_{\Sigma} \, : \, \Sigma
    \ar[
      rr,
      "{
        \scalebox{.7}{
          \color{darkgreen}
          \bf
         $\Sigma$-shaped plot
        }
      }"
    ]
    &&
    \mathcal{X} \,  \text{''}  
    \,.
  \end{tikzcd}
\end{equation}
For instance, if $C=\mathrm{Top}$ and $\Sigma=S^1$ is the circle then $\CX(S^1)$ is interpreted as the \textit{continuous circle}-shaped plots in $\CX$. If instead $C=\CartesianSpaces$ and $\Sigma = \FR^1$ then $\CX(\FR^1)$ is the \textit{smooth path}-shaped plots in $\CX$. If further, $C=\SuperCartesianSpaces$ and
$\Sigma = \FR^{0\vert 1}$ is the fermionic line, or better the \textit{odd infinitesimal point}, then $\CX(\FR^{0\vert 1})$ is the fermionic lines, or \textit{odd points} in $\CX$.

\item By assumption, however, we already have a well-understood notion of morphisms $\mathrm{Hom}_{C}(\Sigma_1,\Sigma_2)$ between any two of our probe spaces in $C$. For this to be consistent with our above interpretation of plots of $\CX$ as arrows valued in $\mathcal{X}$, there should be a way to ``compose'' any
\begin{equation*}
  \label{SchematicsOfProbes}
  \begin{tikzcd}[column sep=large]
     f \, : \, \Sigma'
    \ar[
      rr,
      "{
        \scalebox{.7}{
          \color{darkgreen}
          \bf
         morphism in C
        }
      }"
    ]
    &&
    \Sigma 
    \,.
  \end{tikzcd}
\end{equation*}
with any $\Sigma$-shaped plot $\phi_\Sigma : \Sigma \rightarrow \CX $ to produce a new $\Sigma'$-shaped plot 
\begin{equation}
\label{SchematicsOfPlotComposition}
      \begin{tikzcd}[column sep=large]
       `` \, \Sigma'
        \ar[
          rrr,
          "{ f }",
          "{
            \scalebox{.7}{
              map of probes
            }
          }"{swap, yshift=-2pt}
        ]
        \ar[
          rrrrr,
          rounded corners,
          to path={
               ([yshift=-00pt]\tikztostart.south)
            -- ([yshift=-17pt]\tikztostart.south)
            -- node[yshift=-7pt] {
                  \scalebox{.7}{
                    composite $\Sigma'$-shaped plot
                  }  
               }
               node[yshift=6pt]{
                 \scalebox{.7}{
                   $ f^\ast \phi_\Sigma$
                 }
               }
               ([yshift=-17pt]\tikztotarget.south)
            -- ([yshift=-00pt]\tikztotarget.south)
          }
        ]
        &&&
        \Sigma
        \ar[
          rr,
          "{
            \phi_\Sigma
          }",
          "{
            \scalebox{.7}{
              $\Sigma$-plot
            }
          }"{swap, yshift=-2pt}
        ]
        &&
        \CX \, \text{''} \, .
      \end{tikzcd}
\end{equation}
At this point, one might naively think that for this to make sense, $\CX$ ought to be an object in our original probe category $C$, i.e., a plain probe space. Of course this is undesirable towards our goal of describing more general spaces, and  crucially, this is \textit{not the case}! Indeed, all the above schematic picture requires at a rigorous level is that: For any map of probes $f: \Sigma'\rightarrow \Sigma$ there exists a corresponding ``pullback'' map $f^*_\CX$ of sets which sends $\Sigma$-plots to $\Sigma'$-plots of $\CX$
$$
f^*_\CX \, : \, \ \mathrm{Plots}(\Sigma, \,\CX) \longrightarrow \mathrm{Plots}(\Sigma', \,\CX) \, ,
$$
or equivalently denoted in the notation of \eqref{SetOfPlots} as
\begin{align}\label{PullbackOfPlots}
\CX(f) \equiv f^* _\CX \quad :  \quad   \CX(\Sigma) \longrightarrow \CX(\Sigma') \, ,  
\end{align}
which respects the identity and associativity of composition in $C$, i.e.,
$$
\CX(\id_\Sigma) \, = \, \id_{\CX(\Sigma)} \quad \quad  \mathrm{and} \quad \quad \CX(f\circ g) \, = \, \CX(g) \circ \CX(f)  
$$
for all probes $\Sigma, \Sigma', \Sigma'' \in C$ and all maps of probes $g: \Sigma'' \longrightarrow \Sigma'$, $f : \Sigma' \longrightarrow \Sigma $.

\end{itemize}
To summarize, what we have described so far as our would-be generalized space $\CX$ is simply a  system of $\Sigma$-shaped plots for each probe space $\Sigma \in C$, together with ``pullback'' maps for every map between of probes such that the arrow/plot interpretation from Eqs. \eqref{SchematicsOfProbes} and \eqref{SchematicsOfPlotComposition} is consistent. In mathematical terms, this is nothing but the data defining a (contravariant) \textit{functor} between categories\footnote{The superscript ``$\mathrm{op}$'' here means opposite, encoding the fact that morphisms $f:\Sigma \rightarrow \Sigma'$ change direction under $\CX$ hence ``pulling back'' plots.} of probes and sets 
\begin{align}\label{SpaceAsPreSheaf}
\CX \, : \, C^{\mathrm{op}} \longrightarrow \mathrm{Set} \, .
\end{align}
Following the nomenclature from similar functors arising in topology\footnote{On a fixed topological space (or manifold) M, presheaves $\CX$ in the traditional sense are identified with functors $\CX : \mathrm{Open}(M) \rightarrow \mathrm{Set}$ with the source category being that of open sets with restrictions as maps.}, such a functor is referred to as a \textit{presheaf of sets on the category $C$} and the collection of such presheaves is denoted as
$$
\mathrm{PreSh}(C)\, .
$$

\newpage
\noindent {\bf Glueing via the sheaf condition}

There is one final familiar property that naturally arises in field-theoretic spaces of interest, and also in different fields of mathematics (e.g. topology, algebraic geometry etc.), which we shall require of our notion of spaces: This is a \textit{glueing} condition formally known as the ``\textit{sheaf}'' condition. We shall not need to give the fully general and technical definition for arbitrary presheaves $\CX$ over arbitrary probe categories $C$, as the condition for the actual probe categories relevant to field theory is relatively simple instead (cf. Def. \ref{SmoothSetsDefinition} onwards). Nevertheless, the intuition behind this condition is that given a collection 
$$
\big\{ \phi_{\Sigma_i}  \, \in \,  \CX(\Sigma_i) \big\}_{i\in I}
$$ 
of plots over ``smaller'' probe spaces $\{\Sigma_i\}_{i\in I}$, which happen to cover a ``larger'' probe space $\Sigma$ such that they ``agree on overlaps'', then we should be able to  \textit{glue} these smaller plots to a \textit{unique} larger plot 
$$
\phi_\Sigma \in \CX(\Sigma)\, .
$$
For this to make sense formally, one has to carefully define a consistent notion of ``\textit{coverage}'' of each probe $\Sigma$ by families $\{f_i: \Sigma_i \rightarrow \Sigma\}_{i\in I}$ of maps from other probes in $C$ (see e.g. \cite{Johnstone02}\cite{MacLaneMoerdijk94}). A category $C$ supplied with a coverage is known as a ``\textit{site}''. The sheaf condition on a presheaf $\CX$
is then the demand that for any such cover, the corresponding pullback (restriction) map is injective
$$
\big(\CX(f_i)\big)_{i\in I} \, : \,  \CX(\Sigma) \longrightarrow \coprod_{i\in I} \CX(\Sigma_i)
$$
and surjective on those families that agree on ``intersections'' (assumed to be included as probes via the definition coverage). 

We thus reach our definition of generalized spaces probe-able by a category of simpler test spaces $C$, as those presheaves on $C$ whose plots satisfy the prescribed glueing condition, i.e., as \textit{sheaves} on $C$
$$
\CX \quad \in  \quad \mathrm{Sh}(C) \hookrightarrow \mathrm{PreSh}(C) \, .
$$

\noindent {\bf Maps of generalized spaces as natural transformations }

Having defined our notion of generalized spaces, it remains to identify the correct notion of maps between any two such spaces $\CX$ and $\mathcal{Y}$. As usual in mathematics, this notion of a map should \textit{preserve} the ``internal structure'' of the objects, and as such naturally follows our intuitive discussion above. Indeed, since our spaces are completely determined by their consistent system of $\Sigma$-shaped plots, for each $\Sigma\in C$, the only possible data that a map
$$
h: \CX \longrightarrow \mathcal{Y}
$$
may involve is a corresponding system of maps of sets between $\Sigma$-plots of $\CX$
and $\Sigma$-plots of $\mathcal{Y}$
$$
h_\Sigma: \CX(\Sigma) \longrightarrow \mathcal{Y}(\Sigma)\, .
$$
The only condition is that such a system must be consistent with respect to pullback of probes, in that pulling back along a map $f: \Sigma' \rightarrow \Sigma$ and then applying $h_{\Sigma'}$ is precisely the same as first applying $h_{\Sigma}$ and then pulling back along $f$
\begin{equation}\label{NaturalityOfNatTransf}
h_{\Sigma'} \circ \CX(f) \, = \,  \CX(f) \circ h_\Sigma  \quad : \quad \mathcal{X}(\Sigma) \longrightarrow \mathcal{Y}(\Sigma') \, .
\end{equation}
In other words, the diagram 
\[ 
\xymatrix@C=4em@R=.8em  {\CX(\Sigma) \ar[dd]^{\CX(f)} \ar[rr]^{h_\Sigma} &   &  \CY(\Sigma) \ar[dd]^{\CY(f)}
	\\  \\ 
\CX(\Sigma') \ar[rr]^{h_{\Sigma'}} &  &  \CY(\Sigma') \, 
}    
\]
should commute for all $\Sigma,\Sigma' \in C $ and $f:\Sigma' \rightarrow \Sigma$. In mathematical terms, by following the plot-wise intuition we have arrived at nothing but the definition of a \textit{natural transformation} between the functors $\CX$ and $\mathcal{Y}$,
\begin{align}\label{MorphismsOfSheavesAsNatTrans}
\mathrm{Hom}_{\mathrm{Sh}(C)}(\CX,
\, \CY) \, := \, \mathrm{Nat}(\CX, \, \CY)\, .
\end{align}

Summarizing, by merely attempting to consistently follow the idea that spaces should be determined by the ways one can probe them with simple test spaces of $C$, satisfying a certain gluing condition on their plots, we have arrived at a definition of a whole category of such generalized spaces being sheaves on $C$
$$
\mathrm{Sh}(C) \hookrightarrow  \mathrm{PreSh}(C) \, , 
$$
a (full) subcategory of presheaves on C. Categories of sheaves, i.e., generalized spaces from our perspective, enjoy a lot of positive categorical properties (e.g. limits, colimits and mapping spaces exist etc.), which is indeed what makes them so useful in their mathematical physics applications. The study of the abstract properties of such categories goes under the name \textit{(sheaf) topos theory}, with any category equivalent to a sheaf category being called a \textit{topos}\footnote{Up to a few technical properties required from the category $C$ and its coverage that we need not go into here. The reader interested into the full details and technicalities of topos theory may consult \cite{Johnstone02}\cite{MacLaneMoerdijk94}.}.

\medskip 
\noindent {\bf{Probes as generalized spaces and Consistency of plot-interpretation} }

So far we advocated viewing the sets $\CX(\Sigma)$ defining a (pre)sheaf $\CX$ as the ways to plot $\Sigma$ inside $\CX$, and schematically represented these as arrows $``\phi_\Sigma : \Sigma \rightarrow \CX \text{''}$. Having bootstrapped our definition of generalized spaces $\CX$ as such, it turns out that this picture actually becomes rigorous -- after canonically identifying each of our probes $\Sigma$ with their generalized space avatars $y(\Sigma)$. In more detail, fixing some $\Sigma \in C$ we know what it means to map into it by any other probe $\Sigma' \in C$ using the notion of morphisms in the probe category C. That is, we may define a generalized space $y(\Sigma)$\footnote{It is not in general true that all such ``representable'' presheaves will satisfy the corresponding sheaf condition, for an arbitrary coverage on a category $C$. Nevertheless, this is true in all cases relevant to field theory (one says the corresponding coverages are ``subcanonical''), and we shall assume as such in this exposition.} by declaring its $\Sigma'$-plots to be 
$$
y(\Sigma) ( \Sigma' ) \quad := \quad \mathrm{Hom}_C ( \Sigma', \, \Sigma) \, ,  
$$
with the corresponding pullback maps along probes $g:\Sigma'' \rightarrow \Sigma'$ given canonically by precomposition of maps in $C$
\begin{align*}
y(\Sigma)(g) \, :=  \, g^* \quad : \quad \mathrm{Hom}_C(\Sigma', \, \Sigma) &\longrightarrow \mathrm{Hom}_C(\Sigma'', \, \Sigma)  \\
f &\longmapsto g\circ f \, .
\end{align*}

Having realized any probe $\Sigma$ as a special kind of generalized space 
$$
y(\Sigma) \quad \in \quad \mathrm{Sh}(C)
$$ 
at the same level of any other generalized space $\CX \in \mathrm{Sh}(C)$, there is now an actual notion of mapping from $y(\Sigma)$ into $\CX$. Namely, we may consider the set of morphisms \eqref{MorphismsOfSheavesAsNatTrans} as natural transformations 
\begin{align*}
\mathrm{Hom}_{\mathrm{Sh}(C)}\big(y(\Sigma),
\, \CX\big)
\end{align*}
which are perfectly justified to be considered as ``plots'' of $\Sigma$ into $\CX$. It turns out that this \textit{bona fide} notion of $\Sigma$-plots we have bootstrapped coincides precisely with the \textit{defining} notion of $\Sigma$-plots of $\CX$
$$
\CX(\Sigma)\, .
$$
Indeed, this is nothing but the content of the famous \textit{Yoneda Lemma} in disguise, which yields a canonical bijection (see e.g. \cite[\S 8]{Awodey06})
\begin{align}\label{YonedaLemmaBijection}
\mathrm{Hom}_{\mathrm{Sh}(C)}\big(y(\Sigma), 
\, \CX\big) &\xrightarrow{\qquad \sim \qquad} \CX(\Sigma) \equiv \mathrm{Plots}(\Sigma,\, \CX) \, . \\
h \qquad & \qquad \mapsto \qquad \qquad  h_\Sigma \circ \id_\Sigma  \nonumber
\end{align}
Furthermore, applying this same bijection for the special case where $\CX=y(\Sigma')$ is also a probe space yields  
\begin{align}\label{YonedaLemmaBijectionForRepresentables}
\mathrm{Hom}_{\mathrm{Sh}(C)}\big(y(\Sigma), 
\, y(\Sigma')\big) &\xrightarrow{\qquad \sim \qquad} y(\Sigma')(\Sigma) := \mathrm{Hom}_C(\Sigma,\, \Sigma') \, ,
\end{align}
i.e., the two notions of maps between probes canonically coincide. In other words, no morphisms are lost and no new morphisms arise when probe spaces are viewed as generalized spaces! One says that the category $C$ is \textit{fully faithfully} embedded in its sheaf category $\mathrm{Sh}(C)$ via the Yoneda embedding map 
\begin{equation}\label{YonedaEmbedding}
y\, : \, C  \quad \hookrightarrow \quad \mathrm{Sh}(C)\, . 
\end{equation}

To recap, we have defined our generalized spaces $\CX$ by merely assuming they are fully described by a system $\Sigma$-plots $\CX(\Sigma)$ for some prescribed category of probes $\Sigma\in C$, and so arrived at a category $\mathrm{Sh}(C)$ of all such generalized spaces with a canonical notion of morphisms between them. It just so turns out, however, that the category of probes $C$ is naturally included \eqref{YonedaLemmaBijectionForRepresentables} in that of its generalized spaces, whereby the set of $\Sigma$-plots of $\CX$ is indeed identified \eqref{YonedaLemmaBijection} with the set of \textit{maps} from (the generalized incarnation) of $\Sigma$ into $\CX$. This fully justifies our initial schematic intuition \eqref{SchematicsOfProbes} of viewing $\CX(\Sigma)$ as arrows, or plots, of $\Sigma$ into $\CX$.

By prescribing a different category $C'$ of probe spaces with well-understood structural aspects of different nature as required by the situation at hand, along with an appropriate coverage (glueing prescription), we may define ($\equiv$ probe) the correspondingly structured generalized spaces as its sheaf category $\mathrm{Sh}(C')$. We now move on to show how this works in practice in our physical problems of interest, using the familiar probe categories of Cartesian spaces $\FR^k \in \CartesianSpaces$ for bosonic field theory and then further of super-Cartesian spaces $\FR^{k|q} \in \SuperCartesianSpaces$ when fermions are included. We close by indicating further generalizations\footnote{Apart from these well-established aspects of the physical world, the probe-wise approach to field spaces applies in more speculative theoretical considerations such as that of braided noncommutative field theories \cite{CGRS20}\cite{CGRS21}, by choosing the probe spaces to be a category of appropriate non-commutative nature as indicated in \cite[\S 4.9]{GSz22}. } using probes of infinitesimal disks $\mathbbm{D}^{m}_r$ for the description of infinitesimal/perturbative structure, and abstract simplices $\Delta^k$ for that of higher gauge structure.

\section{Bosonic field spaces and Variational calculus via Smooth Sets}\label{SmoothSetsSec}
Let us briefly recall the textbook prescription of variational calculus for local field theory (see e.g. \cite{Basdevant07}).
\begin{itemize}
\item
The \textit{kinematical data} of a (bosonic) field theory is given by a field (fiber) bundle $\pi : F\rightarrow M$ of smooth manifolds over a $d$-dimensional spacetime $M$\footnote{For simplicity we assume it is oriented. It may be further supplied with extra background structure fields.}. The \textit{(off-shell) field space} is given by sections of the field bundle 
\begin{align*}
\CF \, := \, \Gamma_M(F) = \big\{\phi: M \rightarrow F \, \vert \, \pi \circ \phi = \id_M \big\}\, , 
\end{align*}
which is actually (a priori) 
\textit{only a set} with no further structure. For instance, taking $F$ to be the trivial bundle $M\times N$ recovers the $\sigma$-model field space $C^\infty(M,N)$ with target $N$. Taking instead $F=T^*M$ to be the cotangent bundle recovers the field space of electromagnetism as $U(1)$-gauge potential\footnote{As defined, however, this field space includes only the \textit{trivial topological sector} over $M$. Properly including all sectors together with their gauge transformations defines a groupoid instead (see Sec. \ref{GeneralizationSec}).} 1-forms $\Omega^1_\mathrm{dR}(M)$. 

\item The \textit{dynamics of a field theory} at hand is determined by a \textit{local Lagrangian}
$$
\CL \, : \, \CF \longrightarrow \Omega^{d}(M) \, .
$$
The locality of the Lagrangian means that its value at a field configuration $\phi \in \CF$ may be written in a local chart as a function of the field and its derivatives
$$
\CL(\phi) = L\big(x^\mu, \phi^a,\{\partial_I\phi^a\}_{|I|\leq k}\big) 
= \bar{L}\big(x^\mu, \phi^a, \{\partial_I\phi^a\}_{|I|\leq k} \big) \cdot \dd x^1\cdots \dd x^d \, .
$$
Phrased in global terms, this means that a local Lagrangian is given by the composition 
$$
\CL \, = \, L \circ j^\infty
$$
of the jet prolongation map
$$
j^\infty \, : \, \Gamma_M(F) \longrightarrow \Gamma_M(J^\infty F) \, ,
$$
taking field configurations $\phi$ to the corresponding sections of the infinite jet bundle\footnote{Physical Lagrangians usually factor through a finite order jet bundle $J^n_M F$. However, for the purposes of variational calculus and the variational bicomplex it is convenient to pull these back up to the infinite jet bundle \textit{ab initio} (see e.g.  \cite{Anderson89}\cite{Saunders89}\cite{Zuckerman}\cite{GS25}).},
with a Lagrangian bundle map
\vspace{-2mm} 
\[ 
\xymatrix@C=1.8em@R=.2em  {J^\infty_M F \ar[rd] \ar[rr]^L &   & \wedge^d T^*M \ar[ld]
	\\ 
& M & 
}.  
\]

\item The space of \textit{on-shell fields} 
$$
\CF_{\mathcal{E}\CL}\hookrightarrow \CF
$$
is defined as the set of extrema of the corresponding action 
functional\footnote{Integration over $M$ makes sense only if spacetime is compact, or if the fields have compact support. Generally, one integrates over covering families of compact submanifolds $K_i\hookrightarrow M$ to define a family of action functionals / charges $S_{K_i}$. The extremality condition is then expressed jointly for all such functionals.}
$$
S := \int_M \circ \, \,  \CL \quad : \quad \CF \longrightarrow \FR \, , 
$$
i.e., those field configurations $\phi$ over which arbitrary ``infinitesimal variations'' of the action functional vanish
$$
`` \, \delta S |_\phi \, = \, 0  \, \text{''}\, .
$$
Using the locality of the Lagrangian, the variational principle is the formal integration by parts\footnote{The formal integration manipulation may be identified with a corresponding rigorous algebraic manipulation in the variational bicomplex of the infinite jet bundle $J^\infty_M F$ \cite{Anderson89}\cite{Saunders89}. See also \cite[\S 5]{GS25} for a modern review linking it to our sheaf theoretic context.} to isolate the ``infinitesimal variation'' of the field $\delta \phi$ so that 
\begin{align}\label{NaiveVariationalPrinciple}
\delta S |_{\phi} \, = \, \cdots \, = \, \int_M \langle \CE \CL (\phi) \, , \,  \delta \phi \rangle  \, .  
\end{align}
Here $\CE \CL$ is the corresponding Euler--Lagrange differential operator expressed locally as
\begin{align}\label{EulerLagrangeEquationsLocallyAbusingNotation}
\CE \CL_a (\phi) = \sum_{|I|=0}^{\infty} (-1)^{|I|} \frac{\partial}{\partial x^I} 
\bigg(\frac{\delta \bar{L}\big(x^\mu ,\{\partial_J \phi^b\}_{|J|\leq k} \big)}{\delta (\partial_I \phi^a )} \bigg)\, ,
\end{align}
which upon careful consideration \cite[\S 5]{GS25} is globally identified as a map of sections
$$
\CE \CL = \mathrm{EL}\circ j^\infty \, : \, \CF \longrightarrow \Gamma_M\big(V^*F \otimes \wedge^d T^*M\big)
$$
where $V^*F\rightarrow F$ is the (dual) vertical bundle of $F$ over $M$, and $\mathrm{EL} : J^\infty_M F\rightarrow V^*F \otimes \wedge^d T^*M$ is a bundle map over $F$. Analogously, the ``infinitesimal variation'' $\delta \phi$ is identified as a section 
$$
\delta \phi \quad \in \quad  \Gamma_M(VF) \, 
$$
of the vertical fiber bundle $VF\rightarrow F $ over M, covering the original field $\phi: M \rightarrow F$.
It follows that the natural pairing $\langle - , - \rangle$ is the one induced by the fiber-wise non-degenerate duality pairing of $VF$ with $V^*F$. Thus a field $\phi$ is an extremum or  critical point of the action $S= \int_M \circ \,  \CL$, if and only if it satisfies the \textit{Euler--Langrange equations of motion}\footnote{Here $0_\phi$ denotes the composition of the field $\phi: M\rightarrow F$ with the zero section $0_F: F\rightarrow V^*F\otimes \wedge^d T^*M$ over F.}
$$
\CE \CL (\phi ) \, = \, 0_\phi \quad \in \quad \Gamma_M\big(V^*F \otimes \wedge^d T^*M\big) \, .
$$
\end{itemize}
The task is now to make rigorous sense of the above as ``smooth'' spaces and maps, together with the variational procedure, in a manner which parallels extremality/criticality condition of functions on a smooth finite dimensional manifold $\Sigma$. Recall that for any smooth map $f:\Sigma\rightarrow \FR$ of finite-dimensional manifolds, the usual derivative of the composition with any smooth curve $\gamma_t : \FR^1 \rightarrow \Sigma $ may be written as 
\begin{align}\label{FiniteDimensionVariationFormula}
\partial_t (f\circ \gamma_t)|_{t=0}\, = \, \langle \dd f |_{\gamma} \, , \,  \dot{\gamma}_0\rangle  \, . 
\end{align}
Here $\dot{\gamma}_0 \in T_{\gamma} \Sigma$ is the corresponding tangent vector at the origin of the curve, i.e., an ``infinitesimal variation'' around $\gamma=\gamma_0$, $\dd f : \Sigma \rightarrow T^* \Sigma$ is the differential $1$-form of $f$ and $\langle \, \, , \,\rangle : T^*_{\gamma} \Sigma \times T_{\gamma} \Sigma\rightarrow \FR$ is the canonical pairing between tangent and cotangent vectors. Thus the derivative at $\gamma$ vanishes for all infinitesimal variations, i.e.,  the point $\gamma \in \Sigma$ is an extremum of $f$ if and only if 
\begin{align}\label{FiniteDimensionalCriticality}
\dd f |_{\gamma} \equiv \partial_\mu f(\gamma) \cdot \dd x^\mu \, =  \, 0 \, .
\end{align}
This identifies the critical locus of $f$ with the intersection set of $\dd f$ with the zero-section $0_\Sigma$ inside $T^*\Sigma$, i.e., the ``pullback'' construction
   \vspace{-1mm} 
 \[
\xymatrix@=1.6em  {\mathrm{Crit}(f) \ar[d] \ar[rr] &&   \Sigma \ar[d]^{\dd f} 
	\\ 
	\Sigma \ar[rr]^-{0_\Sigma}  && T^*\Sigma
	\, . } 
\]
  \vspace{-2mm} 

The formulaic analogy of \eqref{FiniteDimensionVariationFormula} and \eqref{FiniteDimensionalCriticality} to the field theoretic case of \eqref{NaiveVariationalPrinciple} and \eqref{EulerLagrangeEquationsLocallyAbusingNotation} is evident, which suggests to identify the space of infinitesimal variations $\Gamma_M(VF)$ with the ``\textit{tangent bundle to field space}''
\begin{align}\label{FieldSpaceTangentBundleSet}
T\CF \, := \, \Gamma_M(VF) \longrightarrow \CF ,
\end{align}
under the projection given by postcomposition of sections with $VF\rightarrow F$, and similarly the space the Euler--Lagrange operator takes values in with the ``\textit{variational cotangent bundle to field space}''
\begin{equation}\label{VariationalCotangentBundleSet}
T^*_\mathrm{var} \CF \, := \, \Gamma_M\big(V^*F \otimes \wedge^d T^*M\big) \longrightarrow \CF \, . 
\end{equation}
This yields the space of on-shell fields $\CF_{\CE \CL}$, i.e.,  the critical locus of the action functional,  as the intersection set of $\CE \CL $ with the zero-section $0_{\CF}$ inside $T^*_\mathrm{var} \CF$
\vspace{-1mm} 
\begin{equation}\label{IntersectionDiagram}
\begin{gathered}
\xymatrix@=1.6em  {\CF_{\CE \CL} \equiv \mathrm{Crit}(S) \ar[d] \ar[rr] &&   \CF \ar[d]^{\CE \CL} 
	\\ 
	\CF \ar[rr]^-{0_\CF}  && T^*_\mathrm{var} \CF
	\, . }
\end{gathered}
\end{equation}
For this field theoretic analogy to rigorously follow as with the finite-dimensional line of reasoning, we need the following requirements.

{\bf Requirements of generalized smooth spaces from bosonic field theory}

\begin{itemize}
  \item[{\bf (i)}] Sections of bundles such as the field space $\CF=\Gamma_M(F)$, top-forms $\Omega^d(M)= \Gamma_M(\wedge^d T^*M)$ or those defining the variational cotangent bundle $T^*_\mathrm{var} \CF = \Gamma_M(V^*F \otimes \wedge^d T^*M)$  should all have a natural smooth structure. Furthermore, the infinite jet bundle $J^\infty_M F$ itself and its space of sections $\Gamma_M (J^\infty F)$ should be smooth spaces of the same nature, i.e., live in the same category of generalized smooth spaces as with the former.

  \item[{\bf (ii)}] The Lagrangian bundle maps $L: J^\infty_M F\rightarrow \wedge^d T^*M$, jet prolongation  $j^\infty: \Gamma_M(F) \rightarrow \Gamma_M(J^\infty F)$ and integration maps $\int_M : \Omega^d(M)\rightarrow \FR$ should preserve the corresponding smooth structures, i.e., should consistute smooth maps in this category of generalized smooth spaces.
  In particular this will imply that the composed local Lagrangian $\CL = L \circ j^\infty$ and action functional $S= \int_M \circ \, \CL$ are also smooth maps.

  \item[{\bf(iii)}] The smooth structure should encode the appropriate notion of a smooth path of fields 
  $$
  \phi_t \, : \, \FR^1 \longrightarrow \CF \, ,
  $$
  so that in particular the composition
  $$
  S\circ \phi_t \, : \, \FR \longrightarrow \CF \longrightarrow \FR
  $$
  is a smooth map in the usual sense of analysis. Notice this assumes that the real line and further Cartesian spaces $\mathbb{R}^k$ should also be viewed as such generalized smooth spaces themselves. The infinitesimal variation of the action functional at $\phi=\phi_0$ should then be \textit{rigorously defined} via the usual derivative, and the variational principle should be \textit{rigorously derived} via the former, so that
$$
\partial_t (S\circ \phi_t) |_{t=0} \, = \, \cdots \, = \,  \int_M \langle \CE \CL (\phi) \, , \,  \dot{\phi}_0 \rangle   \, \, ,
$$
where $\dot{\phi}_0 = \partial_t \phi_t |_{t=0} \in T_\phi \CF \hookrightarrow \Gamma_M(VF)$ defines the corresponding tangent vector over $\phi \in \CF$, i.e., an infinitesimal variation at $\phi$.

\item[{\bf (iv)}] Furthermore, motivated by the study of on-shell constructions and observables (conserved currents, charges etc.), the resulting space of on-shell fields $\CF_{\CE \CL}$ should inherit a natural smooth subspace structure, in that the intersection (pullback, limit) construction \eqref{IntersectionDiagram} exists in the category of these generalized smooth spaces.
\end{itemize}

At this point we invoke our discussion of generalized spaces as being probe-able by simpler test spaces from Sec. \ref{Introduction}. Since we want to encode the notion of smoothness, we take our category of (smooth) probes to be the smooth Cartesian spaces $\FR^k$,
$$
\CartesianSpaces \hookrightarrow \SmoothManifolds\, ,
$$
with the coverage to be that of (differentiably) good open covers. In detail, such a covering of any $\FR^k \in \CartesianSpaces$ is given by a family of maps 
\begin{align}\label{GoodOpenCoverage}
\big\{\iota_i\, : \, \FR^k_i \hookrightarrow \FR^k\big\}_{i\in I}
\end{align} 
from Cartesian spaces of the same dimension, with the property that they are diffeomorphisms onto their open image $U_i := \iota_i (\FR^k_i) \subset \FR^k$ and such that the intersection of any two images $U_{ij}= U_i\cap U_j$ are in turn diffeomorphic to $\mathbbm{R}^k$ (or empty).

This immediately yields our definition of generalized smooth spaces.
\begin{definition}[\bf Smooth Sets]\label{SmoothSetsDefinition}
The category of \textit{smooth sets} is the category of sheaves over Cartesian Spaces 
$$
\SmoothSets \quad  := \quad \mathrm{Sh}(\CartesianSpaces) \, ,
$$
with respect to the (differentiably) good open coverage.
\end{definition}
By the very definition of the good open coverage \eqref{GoodOpenCoverage}, it is not hard to see that a generalized smooth space
$$
\CX \, : \, \CartesianSpaces^{\mathrm{op}} \longrightarrow \mathrm{Set}
$$
satisfies the corresponding sheaf condition, i.e., is a smooth set, if and only if
$$
\CX(\FR^k)
$$
is a sheaf on the manifold $\FR^k$ in the usual sense of topology, for each $\FR^k\in \CartesianSpaces$.

\begin{example}[\bf Manifolds as smooth sets]\label{ManifoldsAsSmoothSets}
Any smooth manifold $\Sigma \in \SmoothManifolds$, such as for instance the real line $\FR$, a spacetime $M$ or the total space of a field bundle $F$, may be viewed as a smooth set via the Yoneda embedding \eqref{YonedaEmbedding} by setting its smooth $\FR^k$-plots to be the set of smooth maps into $\Sigma$,
$$
y(\Sigma) (\FR^k) \quad := \quad \mathrm{Hom}_{\SmoothManifolds}(\FR^k\, , \, \Sigma) \, .
$$
\end{example}
More precisely this is the \textit{restricted} -- but still fully faithful -- Yoneda embedding along \\ $\iota : \CartesianSpaces\hookrightarrow \SmoothManifolds$. What makes this work is the fact that any manifold admits (by definition) a differentiably good open cover by Cartesian spaces. This yields an equivalence between the sheaf categories $\iota^* : \mathrm{Sh}(\SmoothManifolds) \xrightarrow{\sim} \mathrm{Sh}(\CartesianSpaces).$ Intuitively, this is the statement that to know a $\Sigma$-plot of $\CX$ for an arbitrary probe-manifold $\Sigma$ is equivalent to knowing its value along a cover of probe-chart restrictions $\psi : \mathbb{R}^k \hookrightarrow \Sigma$ (by glueing these in the traditional sense).

\begin{example}[\bf Infinite jet bundles as smooth sets]\label{InfiniteJetBundlesAsSmoothSets}
Since any finite order jet bundle $J^n_M F$ is a smooth manifold, we may consider its avatar as a smooth set $y(J^n_M F)  \, \in \,  \SmoothSets$ as per the previous example.
We may then define the limit\footnote{Abstractly, this is the statement that limits in any sheaf category are computed probe-wise via limits in Set.} of these smooth sets as $n\rightarrow \infty$
$$
  y(J^\infty_M F)
  \quad :=
  \quad \mathrm{lim}_n^{\SmoothSets} y(J^n_M F)
  \, ,
$$ 
which amounts to prescribing the smooth $\FR^k$-plots of $y(J^\infty_M F)$ to be  families of $\FR^k$-plots into each $J^n_M F$ ($\equiv$ maps of smooth manifolds)
\begin{align}\label{PlotsOfInfiniteJetBundle}
y(J^\infty_M F)(\FR^k) \quad \cong \quad \Big\{ \big\{s^k_n:\FR^k\rightarrow J^n_M F\;\;  \big|  \;\;   \pi^{n}_{n-1}\circ  s^k_{n}
= s^k_{n-1}\big\}_{n\in \mathbbm{N}} \Big\}\, ,
\end{align} 
which are compatible along the projections $\pi^n_{n-1} : J^{n}_M F \rightarrow J^{n-1}_M F$. In particular an infinity jet $j^\infty_p \phi\in J^\infty_M F$, i.e., a $*$-plot in $y(J^\infty_M F)$, is equivalently represented by the 
compatible family of $n$-jets $\{j^n_p \phi=\pi_n(j^\infty_p)\}_{n\in \mathbb{N}}$, as expected by the underlying set-theoretic limit. 

Pullback of plots \eqref{PullbackOfPlots} along a map of probes $f : \FR^k\rightarrow \FR^{k'}$ for the infinite jet bundle is defined by pulling back each member of the defining family \eqref{PlotsOfInfiniteJetBundle} of an $\FR^k$-plot $s^k_\infty \in y(J^\infty_M F)(\FR^k)$. Lastly, note that the infinite jet bundle this comes with obvious plot-wise projections for each $n\in \mathbbm{N}$ 
\begin{align}\label{InfiniteJetProjections}
\pi^\infty_n \, : \, y(J^\infty_M F) \longrightarrow y(J^n_M F)  \, ,
\end{align}
which is then manifestly compatible with pullback along probes \eqref{NaturalityOfNatTransf}, thus defining a (smooth) map of smooth sets.
\end{example}
By the Yoneda embedding \eqref{YonedaEmbedding}, it follows that traditional real-valued smooth functions $f:J^n_M F \rightarrow N$ on each finite order jet bundle  are uniquely identified with maps of smooth sets
$$ 
f_* \, : \, y(J^n_M F) \longrightarrow y(N) \, ,
$$
all of which may be pulled back along the projection \eqref{InfiniteJetProjections}
to identify the \textit{set of globally finite order functions} on $J^\infty_M F$ \cite{Anderson89}\cite{Saunders89} as a subset of (smooth) maps between smooth sets 
\begin{align}\label{GloballyFiniteOrderFunctions}
C^\infty_{\mathrm{glb}}(J^\infty_M F\, , \, N) \quad \hookrightarrow \quad \mathrm{Hom}_{\SmoothSets}\big(y(J^\infty_M F)\, , \, y(N) \big) 
\end{align}
Carefully carrying this identification through \cite[\S 3-5]{GS25}, the variational bi-complex $\Omega^{\bullet,\bullet}(J^\infty_M F)$ and the corresponding results from \cite{Anderson89}\cite{Saunders89} may be rigorously interpreted as taking place in the bona-fide category of generalized smooth spaces of smooth sets.

\begin{example}[\bf Field spaces as smooth sets]\label{FieldSpacesAsSmoothSets}
Let us now show how a set of sections of a fiber bundle, such as a field space with underlying set of off-shell field configurations $\Gamma_M(F)$ determines a corresponding smooth set. We are after an assignment 
$$
\FR^k \longmapsto \CF(\FR^k)
$$
of \textit{smoothly} $\FR^k$-shaped plots in our field space. But we do have an intuitive notion of smoothly $\FR^k$-parametrized field configurations, namely $\FR^k$-parametrized sections of the field bundle. Thus we set
\begin{align}\label{FieldSpacePlots}
	\CF(\FR^k)\quad :=\quad \{\phi^k:\FR^k\times M \rightarrow F \; | \; \pi\circ \phi^k = \mathrm{pr}_2 \}\, ,
\end{align}

\vspace{-1mm} 
\noindent where $\FR^k\in \mathrm{CartSp}$ and $\mathrm{pr}_2 :\FR^k\times M\rightarrow M$ is the projection onto M. In other words, we take $\FR^k$-plots of fields to be smooth maps of manifolds
$\phi^k: \FR^k\times M \rightarrow F$ such that
	\[ 
\xymatrix@=1.2em{ &&  F \ar[d]^{\pi}
	\\ 
	\FR^k\times M \ar[rru]^-{\phi^k} \ar[rr]^-{\mathrm{pr}_2} && M
}   
\]
\noindent commutes. For example, the point $*$-plots of $\CF$ encode  the set of off-shell field configurations
$\CF(*) \, = \, \Gamma_M (F)$, while the line $\FR^1$-plots of $\CF$ encode the notion of smoothly $\FR^1$-parametrized family of field configurations. 

As one can easily guess, for any map of probes $f:\FR^{k'}\rightarrow \FR^k $ the corresponding pullback of plots is given by
\begin{align*}
\CF(f) \, :=  \, (f\times \id_M)^* \quad : \quad  \CF(\FR^k) &\longrightarrow \CF(\FR^{k'}) \\
\phi^k &\longmapsto \phi^k \circ (f\times \id_M )
\end{align*}
Lastly, it is easy to see that such presheaves $\CF$ satisfy the glueing sheaf condition on each probe $\FR^k$, hence defining actual smooth sets. Note also, in the particular case of a trivial fiber bundle $M\times N$ corresponding to a $\sigma$-model field space, then $\FR^k$-plots of fields reduce to smoothly $\FR^k$-parametrized maps from $M$ into $N$, i.e.
\begin{equation}\label{PlotsOfSigmaModelFieldSpace}
\FR^k\times M \xrightarrow{\quad \phi^k \quad } N \, .  
\end{equation}  
\end{example}

Of course, the above example does not only apply to sections of the field bundle, but to sections of \textit{any} fiber bundles over $M$, such as that of top forms $\wedge^d T^*M$ and the induced bundles $VF$ and $V^*F \otimes \wedge^d T^*M$. That is, the same construction yields the corresponding smooth sets of the top-forms on $M$, tangent bundle \eqref{FieldSpaceTangentBundleSet} and variational cotangent bundle \eqref{VariationalCotangentBundleSet} on $\CF$ 
$$
\underline{\Omega^d}(M), \, T\CF \, , T^* \CF \quad \in \quad \SmoothSets\, .
$$

Moreover, it applies almost verbatim to the case of sections of the infinite jet bundle $\Gamma_M(J^\infty F)$, yielding the corresponding smooth set of sections
\begin{align}\label{SectionsOfJetBundleSmoothSet}
\CF^\infty \quad \in \quad \SmoothSets
\end{align}
Namely, the $\FR^k$-plots of $\CF^\infty$
$$
\CF^\infty (\FR^k)
$$
must be considered as $\FR^k$-parametrized \textit{sections of bundles smooth sets}, i.e. maps $\widetilde{\phi}^k : y(\FR^k \times M)\rightarrow y(J^\infty_M F)$ such that the diagram
	\[ 
\xymatrix@=1.2em{ && y(J^\infty_M F) \ar[d]^{\pi}
	\\ 
	y(\FR^k\times M) \ar[rru]^-{\widetilde{\phi}^k} \ar[rr]^-{\mathrm{pr}_2} && y(M)
}   
\]
commutes. In turn, by the Yoneda embedding \eqref{YonedaEmbedding} and the family-wise plot definition of the infinite jet bundle (Ex. \ref{InfiniteJetBundlesAsSmoothSets}), such maps may be represented by $\FR^k$-parametrized families of sections of finite order jet bundles 
$$
\widetilde{\phi}^k \quad \equiv \quad  \big\{ \widetilde{\phi}^k_n :\FR^k \times M \rightarrow J^n_M F\;\;  \big|  \;\; \pi_M \circ \widetilde{\phi}^k_n  = \mathrm{pr}_2 \, \, \text{and} \, \,    \pi^{n}_{n-1}\circ  \widetilde{\phi}^k_{n}
= \widetilde{\phi}^k_{n-1}\big\}_{n\in \mathbbm{N}}
$$
compatible along the projections $\pi^n_{n-1} \, : \, J^n_M F \rightarrow J^{n-1}_M F$.
\begin{remark}[\bf Internal Hom mapping space construction]\label{InternalHomMappingSpaceConstruction}
It is a fact that any (pre)sheaf category $\mathrm{Sh}(C)\hookrightarrow \mathrm{PreSh}(C)$ enjoys a canonical mapping space construction that goes under the name ``\textit{Internal Hom functor}'' (see e.g. \cite{MacLaneMoerdijk94}). This is an operation that yields a generalized space (of mappings) between any two given generalized spaces $\CX,\CY \in \mathrm{Sh}(C)$
\begin{align*}
[\CX,\, \CY ] \quad \in \quad \mathrm{Sh}(C) \, .
\end{align*}
When this abstract categorical construction is applied in the case of field theoretic spaces of sections within $\SmoothSets$ (see \cite[\S 2]{GS25}), it recovers precisely the smooth sets defined via intuition in Ex. \ref{FieldSpacesAsSmoothSets} .
\end{remark}

\begin{remark}[\bf Non-trivial smooth sets with one point]\label{NonTrivialSmoothSetsWithOnePoint} As an aside,
let us briefly mention here that the definition of smooth sets (Def. \ref{SmoothSetsDefinition}) allows for quite wild spaces. Indeed, there are important spaces that have only one underling point, but nevertheless an infinite amount of higher dimensional $\FR^k$-plots. A striking example is the \textit{classifying} or \textit{moduli space of differential $n$-forms} $\mathbf{\Omega}_\mathrm{dR}^{n}$, defined by assigning its $\FR^k$-plots to be the corresponding set of differential $n$-forms on each Cartesian space $\FR^k$
\begin{align*}
\mathbf{\Omega}_\mathrm{dR}^{n} \, : \, \CartesianSpaces^{\mathrm{op}} &\longrightarrow \mathrm{Set} \\
\FR^k &\longmapsto \Omega^n_\mathrm{dR}(\FR^k) \, .
\end{align*}
For any $0\leq k< n$, its $\FR^k$-plots consist of a single element, the vanishing $n$-form on $\FR^k$. Nevertheless for $k\geq  n$, the $\FR^k$-plots comprise an infinite set! The crucially useful property of this peculiar smooth set is that under the (restricted) Yoneda embedding \eqref{YonedaEmbedding}, the set of maps from a smooth manifold $M$ into $\mathbf{\Omega}_{\mathrm{dR}^n}$ is in canonical bijection the differential n-forms on $M$ (see e.g. \cite[\S 2.3]{GS25})
$$
\big\{\, y(M) \xrightarrow{\quad \quad} \mathbf{\Omega}_{\mathrm{dR}^n} \,  \big\} \quad \cong \quad 
\Omega_{\mathrm{dR}}^n(M) \, .
$$
\end{remark}

Let us now pause and note that with the definition of smooth sets (Def. \ref{SmoothSetsDefinition}) and Examples (\ref{ManifoldsAsSmoothSets}), (\ref{InfiniteJetBundlesAsSmoothSets}) and (\ref{FieldSpacesAsSmoothSets}), we have already accomplished the requirement ${\bf (i)}$ from p. 12-13. Remarkably, we did so without having to discuss any functional analysis or infinite dimensional manifold theory, and we shall proceed with ${\bf (ii)}$ similarly.

{\bf Local Lagrangians and action functionals are maps of smooth sets}

With the descriptions of $\CF$ \eqref{FieldSpacePlots} and  $\CF^\infty$ \eqref{SectionsOfJetBundleSmoothSet} at hand, it is not hard to see \cite[\S 2]{GS25} that the infinite jet prolongation now extends to a map of smooth sets\footnote{Of course, for this plot-wise assignment and those that follow to qualify as maps of smooth sets, they must be compatible with pullbacks of probes \eqref{NaturalityOfNatTransf}. This is straightforward to check for all assignments we shall define here.}
\begin{align}\label{JetProlongationSmoothMap}
j^\infty \, : \, \CF &\longrightarrow \CF^\infty 
\\
\phi^k &\mapsto j^\infty(\phi^k) \nonumber \, .
\end{align}
Explicitly, $j^\infty\phi^k \in \CF^\infty(\FR^k)$ is defined point-wise as $j^\infty \phi^k (x,u):= j^\infty\big(\iota^*_{u\rightarrow \FR^k}\phi^k \big) (x)$, where $u\in \mathbb{R}^k$ is any point in the probe. That is, the prolongation is applied point-wise with respect to the probe space with the derivatives being taken only along $M$. Following \cite[Lem. 3.11]{GS25}, given then any finite order Lagrangian bundle map $L: J^\infty_M F\rightarrow \wedge^d T^*M$ over $M$ we view it as a map of smooth sets \eqref{GloballyFiniteOrderFunctions}, and compose with the jet prolongation \eqref{JetProlongationSmoothMap} to yield local Lagrangians as maps between the smooth set of fields and the smooth set of top-forms
\begin{align}\label{LocalLagrangianSmoothSetMap}
\CL\, := \, L \circ j^\infty \quad : \quad \CF \longrightarrow \underline{\Omega}^d(M) \, .
\end{align}
Explicitly, and as intuitively expected, it may be checked that this composition is defined by sending any $\FR^k$-parametrized field configuration $\phi^k\in \CF(\FR^k)$ to the $\FR^k$-parametrized top-form on $M$
given by the composition
\vspace{-2mm} 
\[ 
\xymatrix@C=1.8em@R=.4em  {& J^\infty_M F \ar[rd] \ar[rr]^L &   & \wedge^d T^*M \ar[ld]
	\\ 
 \FR^k\times M \ar[ru]^{j^\infty\phi^k} \ar[rr] & & M & 
} \, .  
\]
Said simply, in local coordinates one ``carries along'' the dependence on $u\in \FR^k$ so that
\begin{align*}
\CL\big(\phi^k(x,u)\big) \, &=\,  L\big(x^\mu, \phi^{k,a}(x,u),\{\partial_I\phi^{k,a}(x,u)\}_{|I|\leq k}\big) \\
&= \, \bar{L}\big(x^\mu, \phi^{k,a}(x,u), \{\partial_I\phi^{k,a}(x,u)\}_{|I|\leq k} \big) \, \cdot \, \dd x^1\cdots \dd x^d \, .
\end{align*}

Moreover, integration along $M$\footnote{If $M$ is not compact this definition is taken verbatim with respect to any (oriented) compact $d$-dimensional submanifold instead.} extends to a map of smooth sets 
\begin{align}\label{SmoothIntegrationMap}
\int_M : \;\; \underline{\Omega}^{d}(M) \longrightarrow y(\FR)\, .
\end{align}
Expressed plot-wise, for any smoothly $\FR^k$-parametrized top-form $\omega_{\FR^k}\in \underline{\Omega}^{d}(M)(\FR^k)$, the value 
of the function $\int_M \omega_{\FR^k} \in y(\FR)(\FR^k)\cong C^\infty(\FR^k, \FR) $ is given by  
$$
\bigg(\int_M \omega_{\FR^k}\! \bigg)(u) := \int_M \iota_{u}^* \, \omega_{\FR^k}\, ,
$$
where $\iota_u$  is the inclusion $M\xrightarrow{\sim}\{u\}\times M\subset \FR^k\times M $. 
That is, one integrates along $M$ while keeping the $\FR^k$-dependence fixed. Finally, composing the maps of smooth sets \eqref{LocalLagrangianSmoothSetMap} and \eqref{SmoothIntegrationMap}, we get the smooth set map incarnation of the local action functional
\begin{equation}\label{LocalActionSmoothSetMap}
S \, := \, \int_M \, \circ \, \CL \quad : \quad \CF \longrightarrow y(\FR) \, . 
\end{equation}
We thus have also easily accomplished requirement ${\bf (ii)} $ from p. 12-13. 
Moving ahead, we indicate how the requirement ${\bf(iii)}$ from p. 12-13 is also satisfied, and as a bonus ${\bf(iv)}$ as well. 

{\bf Variations via smooth paths of fields and the smooth set on-shell fields}

First we note that both the smooth real line and the field space are viewed at the same level as smooth sets, following Ex. \ref{ManifoldsAsSmoothSets} and Ex. \ref{FieldSpacesAsSmoothSets}
$$
y(\FR), \, \CF \quad \in \quad \SmoothSets \, .
$$
We can thus rigorously consider the notion of smooth paths in $\CF$ as maps of smooth sets
\begin{align}\label{PathOfFieldsInSmoothSets}
\phi_t \, : \, y(\FR) \longrightarrow \CF \, .
\end{align}
Conveniently though, by the bijection of the Yoneda Lemma \eqref{YonedaLemmaBijection}, such maps correspond exactly to $\FR^1$-plots\footnote{Here we abusively use the same symbol for both incarnations of the map.} of $\CF$ (cf. Ex. \ref{FieldSpacesAsSmoothSets})
$$
\phi_t  \quad \in \quad \CF(\FR^1) \, ,
$$
i.e., what we initially thought of as a smoothly $\FR^1$-parametrized family of fields
	\[ 
\xymatrix@=1.2em{ &&  F \ar[d]^{\pi}
	\\ 
	\FR^1\times M \ar[rru]^-{\phi_t} \ar[rr]^-{\mathrm{pr}_2} && M
}   
\]

However, in its plot-wise smooth set map incarnation \eqref{PathOfFieldsInSmoothSets} we may indeed compose with the smooth local action functional \eqref{LocalActionSmoothSetMap} to yield a map of smooth sets
$$
 S \circ \phi_t \quad : \quad y(\FR) \longrightarrow \CF \longrightarrow y(\FR) \, .
$$
Again under the Yoneda embedding \eqref{YonedaEmbedding}, this corresponds precisely to the smooth function on the real line (in the usual sense) given by the image of the corresponding $\FR^1$-plot under $S$  
$$
S(\phi_t) \quad \in \quad y(\FR)(\FR_t) \cong C^\infty(\FR_t,\, \FR) \, .
$$
At this stage we can compute the derivative with respect to $t\in \FR_t$ in the usual sense, and so define the set of extremal points of the action functional as those field configurations $\phi \in \CF(*)$ where the variation of the functional vanishes
$$
\partial_t S(\phi_t) |_{t=0} \, = \, 0 
$$
for all $\phi_t \in \CF(\FR_t)$ such that $\phi= \phi_0:= \iota_{0\hookrightarrow \FR_t}^* \phi_t \, .$

Crucially, however, we may repeat the above argument (cf. \cite[\S 5.3]{GS25}) by considering instead $1$-parameter families of $\FR^k$-plots of fields
$$
\phi^k_t \, : \, y(\FR^k\times \FR_t) \longrightarrow \CF\, , 
$$
or equivalently $\FR^k\times \FR_t$-plots
$$
\phi_t^k \quad \in \quad \CF(\FR^k\times \FR_t) \, .
$$
This allows to consider variations of $S$ on all of its $\FR^k$-plot assignments, thus yielding a general definition of critical $\FR^k$-plots of fields.
\begin{definition}[\bf  Critical plots of action functional]\label{CriticalRkPoints}
Let $S:\CF \rightarrow y(\FR)$ be a local action functional. The \textit{critical $\FR^k$-plots} of $S$ is the subset of $\FR^k$-plots
\vspace{-3mm} 
\begin{align}
		\mathrm{Crit}(S)(\FR^k):=  \Big\{\phi^k\in \CF(\FR^k) \; \big{|}\; \partial_t S (\phi^k_t)|_{t=0}=0, \hspace{0.3cm}
  \forall \, \phi^k_t\in  \CF_{\phi^k}(\FR^k\times \FR^1_t) \Big\} ,		
\end{align}
where $\CF_{\phi^k}(\FR^k\times \FR^1_t) = \big\{\phi^k_t \in \CF(\FR^k\times \FR^1_t) \, | \, \phi^k_{t=0}= \phi^k \in \CF(\FR^k) \big\}$. 
\end{definition}
\noindent Hence, there is an assignment of sets of $\FR^k$-critical plots
\vspace{-1mm}
$$
\FR^k \longmapsto \mathrm{Crit}(S)(\FR^k)\, ,
$$

\vspace{-2mm}
\noindent  for each $k\in \mathbbm{N}$. If $S$ was an arbitrary abstract map of smooth sets, there is no reason why this assignment should be functorial under maps of probes $\FR^{k'}\rightarrow \FR^k$, i.e., would not necessarily define a smooth set.
However, the content of Prop. 5.31 and Cor. 5.32 of \cite{GS25} is precisely that this does define a smooth set in the case of a \textit{local action} functional\footnote{See Prop 5.39 of \cite{GS25} for the case of non-compact spacetimes.} $S=\int_M \, \circ \, \, \CL$, while also rigorously exhibiting this critical smooth set as nothing but the \textit{smooth set Euler--Lagrange locus}, i.e., the smooth set intersection
\begin{equation}\label{SmoothSetEulerLagrangeLocus}
\begin{gathered}
\xymatrix@=1.6em  {\CF_{\CE \CL} \equiv \mathrm{Crit}(S) \ar[d] \ar[rr] &&   \CF \ar[d]^{\mathcal{EL}} 
	\\ 
	\CF \ar[rr]^-{0_\CF}  && T^*_\mathrm{var} \CF
	\,  }
\end{gathered}
\end{equation}
of the Euler--Lagrange operator as a smooth section of the variational cotangent bundle and the zero section over $\CF$. In other words, this identifies the critical $\FR^k$-plots $\phi^k$ of the action functional $S$ as those plots in $\CF(\FR^k)$ which map to the $0$-section $\FR^k$-plot in $T^*\CF(\FR^k)$ covering $\phi^k$ \footnote{More precisely, this is the $\FR^k$-parametrized section of $V^*F \otimes \wedge^d T^*M$ given by composing $\phi^k: \FR^k\times M\rightarrow F$ with the zero section $0_F: F\rightarrow V^*F\otimes \wedge^d T^*M$. } under the  Euler--Lagrange operator (extended to plots analogously to the local Lagrangian from \eqref{LocalLagrangianSmoothSetMap})
$$
\CE \CL (\phi^k ) \, = \, 0_{\phi^k}  \quad : \quad \FR^k \times M \longrightarrow V^*F \otimes \wedge^d T^*M \, . 
$$
With this result, we have successfully achieved both requiremenets ${\bf (iii)}$ and ${\bf (iv)}$ from p. 12-13.

{\bf Further field theoretic concepts recognized to rigorously take place in smooth sets}

Let us close this overview on smooth sets and bosonic field theory by listing many familiar field theoretic concepts and results that have been naturally defined and proven within the setting of smooth sets in \cite{GS25}. These include, but are not limited to :The notions of finite diffeomorphisms and infinitesimal ones as smooth (local) vector fields on field spaces $\CF$ \cite[\S 2 ]{GS25}; Finite (local) symmetries, currents and charges of Lagrangian field theories \cite[\S 3 ]{GS25}; Infinitesimal (local) symmetries, conserved currents, Noether's 1st and 2nd Theorems, Cauchy surfaces \cite[\S 6]{GS25}; Tangent bundles of on-shell field spaces $T \CF_{\CE \CL}$ ($\equiv$ ``Jacobi fields'')\footnote{Note however, this is not ``naturally'' defined within smooth sets as it involves \textit{actual} infinitesimal curves (cf. Sec. \ref{GeneralizationSec}).}, the local bicomplex on $\CF\times M$ of Zuckerman \cite{Zuckerman} and, last but not least, the canonical presymplectic current $\omega_{\CL}$ of any local Lagrangian field theory $(\CF, \CL)$ hence defining the on-shell \textit{covariant phase space} as a presymplectic smooth set \cite[\S 7]{GS25}.

\section{Fermionic field spaces via Super Smooth Sets}\label{SuperSmoothSetsSec}
We now change gears and go back at an intuitive situation as a source of motivation. Following \cite{Freed99}, we consider one of the simplest examples of a Lagrangian containing a fermionic field; that of the fermionic particle on the real line
\begin{equation}\label{FermionicParticleLagrangian}
\CL_{\mathrm{Fer. Part.}} \, = \, \psi
\, \partial_t \psi \cdot \mathrm{d} t \, . 
\end{equation}
In this expression $\psi$ is manipulated as an ``anticommuting variable'' representing the fermionic field at the classical level. We wish to ask what is the symbol $\psi$  mathematically; which set does it belong to, if any? 

{\bf Naive attempts at describing a fermionic field}
\begin{itemize}
    \item[{\bf(i)}] A first naive guess is that $\psi$ is a simply a real-valued function on the real line. Namely, an element of
    $$
    C^\infty( \FR_t \, , \, \FR) \, .
    $$
    But this fails immediately as in this case the elements $\psi$ and $\partial_t \psi$ would commute, and so the Lagrangian \eqref{FermionicParticleLagrangian} would be trivial, i.e., a total derivative
    $$
    \CL (\psi) = 2 \, \partial_t (\psi^2) \, \dd t \, = 2 \, \dd \psi \, .
    $$

    \item[{\bf (ii)}] In view of the anticommuting nature of $\psi$ in contrast to the commuting nature of the source time-line $\mathbb{R}_t$, another attempt is to consider the target of the maps to be the \textit{odd} real line 
    $$
    \FR^{0\vert 1} \, \equiv \, \FR_\mathrm{odd} \, .
    $$ 
    This is the supermanifold defined by the polynomial algebra of functions generated by one odd (coordinate) variable $b$
    $$
    \CO(\FR^{0\vert 1}) \quad := \quad \FR[b] \quad \cong  \quad \FR\oplus b\cdot \FR \, .
    $$
    Since maps of supermanifolds are completely determined by their \textit{pullback} action on their function super-algebras (cf. discussion around \eqref{SuperCartesianSpaces}), 
    $$
\mathrm{Hom}_{\SuperManifolds}(\FR_t \, , \, \FR^{0\vert 1}) \quad \cong \quad \mathrm{Hom}_{\mathrm{\mathrm{Sup}CAlg}} \big(\CO(\FR^{0\vert 1}) \, , \, C^\infty(\FR_t) \big) \, ,
    $$
any map $\psi : \FR_t \rightarrow \FR^{0\vert 1}$ is completely determined by the value of the generating coordinate $b$ under the pullback morphism of algebras 
\begin{align*}
\psi^* \, : \, \CO(\FR^{0\vert 1}) &\longrightarrow C^\infty(\FR_t)  \, . 
\end{align*}
But since $b$ is an odd element, it must be that $$
\psi^*(b)\cdot \psi^*(b)=\psi^*(b^2)=\psi^*(0)=0 
$$
and the only smooth function $\psi^*(b) \in C^\infty(\FR_t)$ on the real line that squares to zero is the constant 0-function, i.e., necessarily
$$
\psi^*(b) \, = \, 0 \, . $$
In other words,
$$
\mathrm{Hom}_{\SuperManifolds}(\FR_t \, , \, \FR^{0\vert 1}) \quad \cong \quad \{ 0\} 
$$
and $\psi$ would necessarily represent the trivial fermion field!
\item[{\bf (iii)}] One notices, however, that the above line of logic would produce a non-trivial set if we adjoined an ``\textit{auxilliary odd coordinate}'' $\theta$ to the function algebra of the time line
$$
\mathrm{Hom}_{\mathrm{\mathrm{Sup}CAlg}} \big(\CO(\FR^{0\vert 1}) \, , \, C^\infty(\FR_t) [\theta] \big)\, ,
$$
since such a (pullback) map of algebras is then determined by its action on the generator $b\in \CO(\FR^{0\vert 1})$ as
$$
b \longmapsto f(t) \cdot \theta 
$$
for an arbitrary choice of function $f(t)\in C^\infty(\FR_t)$. In other words,
\begin{equation}\label{AlgebraMapsOneAuxiliary}
\mathrm{Hom}_{\mathrm{\mathrm{Sup}CAlg}} \big(\CO(\FR^{0\vert 1}) \, , \, C^\infty(\FR_t)[\theta] \big)
\quad \cong \quad C^\infty(\FR_t) \cdot \theta \, .
\end{equation}
In terms supermanifolds, the algebra with one extra auxiliary coordinate is the algebra of functions on the product $\FR_t\times \FR^{0\vert 1}_\theta$
$$
\CO(\FR_t\times \FR^{0\vert 1}_\theta) \quad := \quad C^\infty(\FR_t)\otimes \CO(\FR^{0\vert 1}_\theta) \quad \cong \quad C^\infty(\FR_t) [\theta]\, ,
$$ 
which by \eqref{AlgebraMapsOneAuxiliary} implies that, dually, via maps 
of supermanifolds
$$
\mathrm{Hom}_{\SuperManifolds}\big(\FR^{0\vert 1}_{\theta}\times \FR_t \,, \, \FR^{0\vert 1} \big) \quad \cong \quad C^\infty(\FR_t) \cdot \theta \, .
$$
This does provide a non-trivial ``set of fields'' of fermions of anticommuting nature, but nevertheless on which the Lagrangian \eqref{FermionicParticleLagrangian} still acts trivially. Indeed for any $\psi_\theta= f(t) \cdot \theta$, one has
$$
\CL (\psi_\theta) = f(t) \cdot \partial_t f(t) \cdot \theta^2 \cdot \dd t = 0\, \quad \in \quad \Omega^1_\mathrm{dR}(\FR_t)\otimes \CO(\FR^{0\vert 1}_\theta)\, ,  
$$
by virtue of $\theta^2 = 0$.

\item[{\bf (iv)}] To circumvent the triviality of the Lagrangian in {\bf (iii)}, we may supply instead 2 auxiliary odd coordinates by considering instead the set of mappings from $\FR^{0\vert 2}_{ \theta_1\theta_2} \times \FR_t \cong \FR^{0\vert 1}_{\theta_1}\times \FR^{0\vert 1}_{\theta_2}\times \FR_t$ into $\FR^{0\vert 1}$. Carrying through the dual algebra calculation as above, it follows then that 
$$
\mathrm{Hom}_{\SuperManifolds}\big(\FR^{0\vert 2}_{\theta_1\theta_2}\times \FR_t \,, \, \FR_\odd \big) \quad \cong \quad C^\infty(\FR_t) \cdot \theta_1 \oplus C^\infty(\FR_t) \cdot \theta_2 \, , 
$$
whose elements are of the form
$$
\psi_{\theta_1 \theta_2} \, = \, f_1(t)\cdot \theta_1 + f_2(t) \cdot \theta_2 \, .
$$
This implies that the Lagrangian \eqref{FermionicParticleLagrangian} does act \textit{non-trivially} as a map
$$
\CL \, : \, \mathrm{Hom}_{\SuperManifolds}\big(\FR^{0\vert 2}_{\theta_1 \theta_2}\times \FR_t \,, \, \FR_\odd \big) \xrightarrow{\quad \quad } \Omega^1_\mathrm{dR}(\FR_t)\otimes\CO(\FR^{0\vert 2}_{\theta_1 \theta_2}) \, , 
$$
since
$$
\CL (\psi_{\theta_1 \theta_2})=\cdots = \big(f_1\cdot \partial_t f_2\, {\color{red}-} \,f_2 \cdot \partial_t f_1 \big )\cdot \theta_1  \theta_2 \cdot \dd t 
$$
is non-zero and moreover \textit{non-exact}, due to the crucial minus sign appearing by anticommuting the auxilliary coordinates $\theta_1$ and $\theta_2$.

\item[{\bf (v)}] Upon 
further inspection, the reason the Lagrangian requires 2 auxiliary odd coordinates to exhibit its non-triviality is because it is a 2nd order polynomial in $\psi$. It follows similarly that to display a fermionic Lagrangian of polynomial order $q\in \mathbbm{N}$ as a non-trivial map, or any observable functional of the fermionic field $\psi$ for that matter, requires the introduction of (at least) $q$ auxiliary odd coordinates so that
\begin{align}\label{LagrangianAtOrderq}
\CL_{\FR^{0\vert q}_\mathrm{aux}} \, : \, \mathrm{Hom}_{\SuperManifolds}\big(\FR^{0\vert q}_{\mathrm{aux}}\times \FR_t \,, \, \FR^{0\vert 1} \big) &\xrightarrow{\quad \quad} \Omega^1_\mathrm{dR}(\FR_t)\otimes\CO(\FR^{0\vert q}_{\mathrm{aux}}) \\
\psi_{\FR^{0|q}_\mathrm{aux}} \quad \quad &\quad \longmapsto \quad \quad  \CL(\psi_{\FR^{0|q}_\mathrm{aux}} ) \nonumber
\end{align}
where $\FR^{0|q}_{\mathrm{aux}}:= \FR^{0|1}_{\theta_1} \times \cdots \times  \FR^{0|1}_{\theta_q}$ is an ``auxiliary'' $q$-dimensional odd Cartesian space. 
\end{itemize}

We have thus argued that the symbol $\psi$ is to be interpreted as a morphism of supermanifolds from the time-line $\FR_t$ to the odd line $\FR^{0\vert 1}$, but \textit{parametrized} by an extra auxilliary odd Cartesian space $\FR^{0|q}$ as necessary by the polynomial order of the Lagrangian (or observable) at hand. This means that $\psi$ is not an element of a fixed set, and hence still begs the question of what is \textit{actually} the fermionic field space it represents?

{\bf Super sets}

The attentive reader will immediately notice that the above pattern is completely analogous to the smooth case, where we considered $\FR^k$-plots of a bosonic field space as $\FR^k$-parametrized smooth maps of manifolds (cf. Eq. \eqref{PlotsOfSigmaModelFieldSpace} of Ex. \ref{FieldSpacesAsSmoothSets}). That is, what we are implicitly describing as $\FR^{0\vert q}$-parametrized maps of supermanifolds in \eqref{LagrangianAtOrderq} are nothing but the \textit{plots} of the corresponding \textit{super space of fermionic fields} traced out by the odd Cartesian space $\FR^{0\vert q}$, in line with the intuition of Sec. \ref{Introduction}. This is indeed encoded in the following notion of generalized space of a \textit{super set} \cite[\S 4.6]{dcct}, as per the discussion of Sec. \ref{Introduction}. To state this properly, we denote by $\mathrm{Odd}\CartesianSpaces$ the category of odd Cartesian probe-spaces, also known as \textit{super points} (see Ex. \ref{OddCartesianSpacesAsSuperPoints}), namely opposite category to that of finite Grassmann algebras generated by a finite number of odd variables
\begin{align}\label{OddCartesianIntoGrassmannOp}
\mathrm{Odd}\CartesianSpaces &\xrightarrow{\quad \sim\quad } \mathrm{Grassmann}_\FR^{\mathrm{op}} \\
\FR^{0\vert q}&\quad \longmapsto \quad \CO(\FR^{0\vert q}) := \FR[\theta_1,\cdots, \theta_q] \, .\nonumber 
\end{align}

\begin{definition}[\bf Super Sets]\label{SuperSets}
The category of \textit{super sets} is the category of sheaves over odd Cartesian Spaces 
$$
\mathrm{SupSets} \quad  := \quad \mathrm{Sh}(\mathrm{Odd}\CartesianSpaces) \, ,
$$
with respect to the trivial coverage $\{ \id_{\FR^{0\vert q}} :\FR^{0\vert q} \rightarrow\FR^{0\vert q} \}$.
\end{definition}
\noindent In particular, considering the trivial coverage implies the glueing (sheaf) condition is satisfied by all presheaves on $\mathrm{Odd}\CartesianSpaces$, hence any functor 
$$
\CX \, : \, \OddCartesianSpaces \longrightarrow \mathrm{Set} 
$$
constitutes a super set.

\begin{example}[\bf Odd Cartesian spaces as super points]\label{OddCartesianSpacesAsSuperPoints}
By the Yoneda embedding \eqref{YonedaEmbedding}, any probe odd Cartesian space $\FR^{0\vert l}$ may be viewed as a super set
$$
y(\FR^{0\vert l}) \quad \in \quad \SuperSets \, .
$$
Notice any such super set has only \textit{one point}, in that its $\FR^{0|0}\equiv*$-plots are given by a single element
\begin{align*}
y\big(\FR^{0|l}\big)(*) :&= \mathrm{Hom}_{\OddCartesianSpaces}\big( * \, , \, \FR^{0\vert l} \big) \, = \, \mathrm{Hom}_{\mathrm{Sup}\mathrm{CAlg}}\big( \CO(\FR^{0\vert l})\, , \, \FR \big) \\ 
& \cong  \{0\} \, ,
\end{align*}
where $\FR\cong \CO(*)$ is the function algebra of the point. Nevertheless, one can check that its odd $\FR^{0|q}$-plots for $q>0$ are potentially non-vanishing so that
$$
y(\FR^{0|l})(\FR^{0|q}) := \mathrm{Hom}_{\OddCartesianSpaces}( \FR^{0\vert q} \, , \, \FR^{0 \vert l} ) \, \neq \{0\} \, .
$$
This is one way to justify the naming of odd Cartesian spaces as the \textit{super points}: Their underlying set of points is a singleton, but nevertheless their infinitesimal fermionic nature can be non-trivially probed with other odd Cartesian spaces\footnote{Of course, this is in line with their locally ringed space definition. As a ringed space $\FR^{0\vert q}=(*\, , \, \FR[\theta_1,\cdots \theta_l])$, i.e., it has underlying topological space the point but supplied with the non-trivial ring (of functions) given by the corresponding Grassmann algebra.}.
\end{example}

\begin{example}[{\bf Fermionic particle field as a super set}]\label{FermionicParticleFieldAsASuperSet}
Let $\FR_t$ be the source real-time line and $\FR^{0\vert 1}$ be the target odd line for the fermionic particle. Then the $\FR^{0\vert q}$-parametrized maps described in \eqref{LagrangianAtOrderq}  define its space of fields as the super set with plots
\begin{align*}
\CF_{\mathrm{Fer. Part.}} \quad : \quad  \OddCartesianSpaces &\longrightarrow \mathrm{Set} \\
\FR^{0\vert q} \quad &\longmapsto \quad \mathrm{Hom}_{\SuperManifolds}\big(\FR^{0\vert q}_{\mathrm{aux}}\times \FR_t \,, \, \FR^{0\vert 1} \big) \, . 
\end{align*}
Moreover the assignment from \eqref{LagrangianAtOrderq} can be checked to be functorial under maps of probe super points $f: \FR^{0\vert q'} \rightarrow \FR^{0\vert q}$, hence exhibiting the Lagrangian formula \eqref{FermionicParticleLagrangian} as a well-defined map of super sets
$$
\CL \quad : \quad \CF_{\mathrm{Fer. Par.}} \longrightarrow \underline{\Omega}^{1}_{\mathrm{dR}}(\FR_t)
$$
where the super set on the right is determined by $\FR^{0\vert q}\mapsto \Omega^{1}_{\mathrm{dR}}(\FR_t)\otimes \CO(\FR^{0\vert q})$. 
\end{example}

This fully incorporates and explains the attempts from p. 20-22. Namely, such a fermionic field space has only one point, i.e., one $*$-plot
$$
\CF_{\mathrm{Fer. Part.}}(*) \quad = \quad \{\psi^{0\vert0} = 0\} \, ,
$$
but nevertheless has non-trivial higher odd-plots (cf. the purely smooth case of Rem. \ref{NonTrivialSmoothSetsWithOnePoint}),
$$
\CF_{\mathrm{Fer. Part.}}(\FR^{0\vert q}) \quad \equiv  \quad  \big\{  \psi^{0\vert q} \, : \, \FR^{0\vert q}_{\mathrm{aux}}\times \FR_t \longrightarrow \FR^{0\vert 1} \big\}\quad \neq \quad \{0\} \, 
$$
for $q> 1$. Moreover, the Lagrangian of the free fermionic particle holds trivial information at the level of both point and $\FR^{0\vert 1}$-plots, but nevertheless
encodes non-trivial information at higher plots,
$$
\CL_{\FR^{0\vert 0}}  (\psi^{0\vert 0}) \, =\,  0  \qquad , \qquad \CL_{\FR^{0\vert 1}}  (\psi^{0\vert 1}) \, =\,  0 \qquad , \qquad \CL_{\FR^{0\vert 2}}  (\psi^{0\vert 2}) \, \neq \,  0 \,  
$$
for $q>1$.

In fact, the argument we made  in p. 20-22 towards the probe-wise description of fermionic field spaces applies verbatim for more general fermionic fields $\psi$, i.e., being sections of an arbitrary \textit{odd vector field bundle}
$$
V_\odd \longrightarrow M
$$
over a spacetime $M$ and any polynomial Lagrangian $\CL$ in $\psi$. The corresponding super set of fermionic fields is then given by the probe-wise assignment of plots
\begin{align}\label{FermionicFieldSpaceOddPlots}
	\CF_{\mathrm{Fer.}}(\FR^{0\vert q})\quad :=\quad \big\{\psi^{0|q}:\FR^{0\vert q}\times M \rightarrow V_\odd \; | \; \pi\circ \psi^{0|q} = \mathrm{pr}_2 \big\} \, ,
\end{align}
where the arrows here mean maps of supermanifolds, i.e., dually pullback morphisms between the corresponding super-algebras of functions (see later Ex. \ref{Boson-FermionFieldSpacesAsSuperSmoothSets} for more details). For instance, this is already necessary to describe mathematically the physical Dirac fermion Lagrangian 
$$
\CL_{\mathrm{Dirac}} \, = \, (\overline{\psi} \gamma^\mu \partial_\mu \psi ) \cdot  \dd^4x \, , 
$$
where $V_\odd$ is here the odd version of the corresponding Spinor bundle, say over Minkowski spacetime $M= \FR^{1,3}$. Such local fermionic Lagrangians, and generally observables, may be checked explicitly to be functorial under maps of probes \eqref{NaturalityOfNatTransf}, hence displaying that the fermionic formulas naively written in the physical literature are implicitly well-defined, and actually representing \textit{morphisms of super sets}! 

So far this fully formalizes and explains the usual description of classical fermionic fields (see e.g. \cite{Freed99}). Note, however, if we are to the formulate the corresponding variational calculus rigorously in analogy to the bosonic case described in Sec. \ref{SmoothSetsSec}, we must also be able to consider smooth paths of ($\FR^{0\vert q}$-plots of) fermionic fields -- that is we would also like to describe the smooth structure of such field spaces. Following \cite[\S 4.6]{dcct}, the smooth and super structure of field spaces may be simultaneously described by enhancing our collection of probes to include both Cartesian, odd Cartesian spaces and their products.

{\bf Super Cartesian spaces and Super Smooth sets}

To properly state the definition of super smooth sets, let us first precisely identify our category of probes. Firstly recall the standard result (``Milnor's exercise'', cf. \cite[\S 35.8-10]{KMS}) that states that any smooth manifold $M$, and in particular any Cartesian space $\FR^k$, is completely determined by its algebra of functions. Namely, the functor that extracts the algebra of functions is a fully faithful embedding into commutative algebras
\begin{align*}
\CartesianSpaces  \quad &\hookrightarrow \quad \mathrm{CAlg}^{\mathrm{op}}_\FR \\
\FR^k \quad &\mapsto \quad C^\infty(\FR^k)\, .
\end{align*}
Since any commutative algebra is a a special case of a super-algebra with no odd elements $\mathrm{CAlg}_\FR \hookrightarrow  \mathrm{Sup}\mathrm{CAlg}_\FR$, Cartesian spaces may be viewed as such via their function algebras
$$
\CartesianSpaces 
 \quad \hookrightarrow \quad \mathrm{Sup}\mathrm{CAlg}^{\mathrm{op}}_\FR \, .
$$

Thinking of super-algebras abstractly as function algebras of would-be generalized supermanifolds, we enlarge our collection of smooth probes $\CartesianSpaces$ and  \textit{define} our category of super smooth probes to be formal duals of algebras of the form $C^\infty(\FR^k)\otimes \CO(\FR^{0\vert q})$, namely
\begin{align}\label{SuperCartesianSpaces}
\SuperCartesianSpaces \quad &\hookrightarrow \quad \mathrm{SupCAlg}^{\mathrm{op}}_\FR \\
\FR^{k\vert q} \equiv \FR^k\times \FR^{0\vert q} \quad &\mapsto \quad \CO(\FR^{k\vert q}) := C^\infty(\FR^k)\otimes \CO(\FR^{0|q})\, , \nonumber
\end{align}
where $\CO(\FR^{0|q}) = \FR[\theta_1,\cdots \theta_q]$ is the algebra of the corresponding super-point. We supply it with the (differentiably) good open coverage, trivially extended along odd directions. In detail, this declares a covering of any $\FR^{k\vert q} \in \SuperCartesianSpaces$ to be given by a family of maps 
\begin{align}\label{SuperGoodOpenCoverage}
\big\{\iota_i \times \id_{\FR^{0\vert q}}\, : \, \FR^k_i \times \FR^{0|q} \hookrightarrow \FR^{k\vert q} \big\}_{i\in I}
\end{align}
where the restricted family $\{\iota_i :  \FR^k_i \hookrightarrow \FR^k \}$ is a (differentiably) good open coverage of Cartesian spaces in the sense of \eqref{GoodOpenCoverage}.

\begin{definition}[\bf Super Smooth Sets]\label{SuperSmoothSetsDefinition}
The category of \textit{super smooth sets} is the category of sheaves over super Cartesian Spaces 
$$
\SuperSmoothSets \quad  := \quad \mathrm{Sh}(\SuperCartesianSpaces) \, ,
$$
with respect to (differentiably) good open coverage extended trivially in the odd directions \eqref{SuperGoodOpenCoverage}.
\end{definition}
Similar to the case of purely smooth sets from Def. \ref{SmoothSetsDefinition}, the coverage \eqref{SuperGoodOpenCoverage} implies that a generalized super smooth space
$$
\CX \, : \, \SuperCartesianSpaces^{\mathrm{op}} \longrightarrow \mathrm{Set}
$$
satisfies the corresponding sheaf condition, i.e., is a super smooth set, if and only if
$$
\CX(\FR^{k\vert q}) \, \equiv \, \CX(\FR^k\times \FR^{0\vert q})
$$
is a sheaf on the manifold $\FR^k$ when in the usual sense of topology when restricted along $\iota : \FR^k \hookrightarrow \FR^k\times \FR^{0\vert q}$, for each $\FR^k\in \CartesianSpaces$ and every $\FR^{0\vert q}\in \OddCartesianSpaces$.

\begin{example}[{\bf Supermanifolds as super smooth sets}]\label{SupermanifoldsAsSuperSmoothSets}

Any super manifold $\Sigma \in \SuperManifolds$, such as a spacetime $M$ or an odd vector bundle $V_\odd \rightarrow M$, may be viewed as a super smooth set via the Yoneda embedding \eqref{YonedaEmbedding} by defining its  $\FR^{k\vert q}$-plots to be the set of maps of supermanifolds into $\Sigma$,
\begin{align*}
y(\Sigma) (\FR^{k\vert q}) \quad :&= \quad \mathrm{Hom}_{\SuperManifolds}(\FR^{k\vert q}\, , \, \Sigma) \\
&\cong \quad \mathrm{Hom}_{\mathrm{SupCAlg}_\FR}\big( \CO(\Sigma) \, , \, C^\infty(\FR^k)\otimes \CO(\FR^{0\vert q}) \big) \, .
\end{align*} 
\end{example}
Here we implicitly use the fact that super manifolds $\Sigma$ (as locally ringed spaces modelled on $\FR^{k\vert q})$, are also fully determined by their super algebra of (global) functions $\CO(\Sigma)$ (e.g., as a corollary of Bachelor's theorem \cite{Batchelor79}). More precisely then, this example employs the \textit{restricted} -- but still fully faithful -- Yoneda embedding along $\iota : \SuperCartesianSpaces\hookrightarrow \SuperManifolds$\footnote{As with the purely smooth case, this works because any super manifold admits (by definition) a differentiably good open cover by super Cartesian spaces. This yields an equivalence between the sheaf categories $\iota^* : \mathrm{Sh}(\SuperManifolds) \xrightarrow{\sim} \mathrm{Sh}(\SuperCartesianSpaces).$}.

\begin{example}[{\bf Boson-Fermion field spaces as super smooth sets}]\label{Boson-FermionFieldSpacesAsSuperSmoothSets}
Consider now an odd vector (fermion) field bundle $V_\odd \rightarrow M$ over spacetime, or more generally a composite
$$
F^{V_\odd} \xrightarrow{\quad \quad } F\xrightarrow{\quad \pi \quad } M \, .
$$
odd vector bundle over $F$, which in turn is a bosonic fiber bundle over $M$. The total space of such a composite is a field bundle encodes the internal nature of both bosonic \textit{and} fermionic fields, simultaneously (cf. \cite{GianchettaGiovanniSardanashvily09}). For instance, in the case where the fields are indendently defined\footnote{This is \textit{not} the case when considering a gravitational theory with coupled fermionic spinorial fields. The latter's definition depends on the choice of metric $g$, hence the fully general composite description is necessary.} $F^{V_\odd} $ could be the pullback of a separate odd vector bundle over $M$ 
$$
F^{V_\odd}  \quad = \quad \pi^*V_\odd \, . 
$$

The \textit{smooth super space of off-shell fields} on $M$ is defined as the super smooth set defined by the assignment
$$
\FR^{k\vert q} \longmapsto \CF(\FR^{p\vert q})
$$
of $\FR^{k\vert q}$-shaped plots, which are given by $\FR^{k\vert q}$-parametrized sections of the composite field bundle. Thus we set
\begin{align}\label{BosonicFermionicFieldSpacePlots}
	\CF(\FR^{k\vert q})\quad :=\quad \big\{\phi^{k\vert q}:\FR^{k\vert q}\times M \rightarrow F^{V_\odd}  \; | \; \pi_M \circ \phi^{k\vert q} = \mathrm{pr}_2 \big \}\, ,
\end{align}

\vspace{-1mm} 
\noindent where $\FR^k\in \mathrm{CartSp}$ and $\mathrm{pr}_2 :\FR^k\times M\rightarrow M$ is the projection onto M. Of course, here the maps are explicitly computed dually in terms of pullbacks of function superalgebras as per Ex. \ref{SupermanifoldsAsSuperSmoothSets}. In other words, these are taken to be maps of super-manifolds
$\phi^{k\vert q}: \FR^{k\vert q}\times M \rightarrow F^{V_\odd} $ such that
	\[ 
\xymatrix@=1.2em{ &&  F^{V_\odd}  \ar[d]^{\pi}
	\\ 
	\FR^{k\vert q}\times M \ar[rru]^-{\phi^{k\vert q}} \ar[rr]^-{\mathrm{pr}_2} && M
}   
\]
\noindent commutes. The functoriality under maps of probes $f : \FR^{k'\vert q'}\rightarrow \FR^{k\vert q}$ and the sheaf condition hold analogously to the purely bosonic case of Ex. \ref{FieldSpacesAsSmoothSets}.
\end{example}

One can readily check that the points of such a field space are given simply by the set of off-shell \textit{bosonic} field configurations
$$
\CF(\FR^{0\vert 0}) \quad \cong \quad \Gamma_M (F)\, ,
$$
extending the fact that fermionic fields do not appear at the point-set level of geometry from Ex. \ref{FermionicParticleFieldAsASuperSet}. The non-trivial fermionic information is encoded solely in higher odd plots of this smooth super set. To simplify the formulas let us consider this for the case of $F^{V_\odd}=\pi^*V_\odd$, so that one may check in particular that
$$
\CF(\FR^{0\vert 1}) \quad \cong \quad \Gamma_M(F)\times \big(\Gamma_M (V) \cdot \theta\big ) \, , 
$$
where $\theta$ is the odd coordinate of the probe space $\FR^{0\vert 1}$. This reveals both the bosonic configurations \textit{and} also non-trivial fermionic configurations, but only for linear Lagrangians and observables, generalizing the case of the fermionic particle from {\bf (iii)} of p. 21. For higher order Lagrangians and observables in the fermion field, higher odd plots are required and computed similarly.

More generally, and in particular towards the purpose of variations of action functionals, 
one may consider smoothly parametrized families of such odd plots of fields. For example, it follows similarly that
$$
\CF(\FR^{1\vert 1})\quad \quad \cong \quad  \Gamma_{\FR^1_t\times M }(\mathrm{pr}_2^* F) \times \big(\Gamma_{\FR^1_t\times M}(\mathrm{pr}_2^* V)\cdot \theta\big) \, , 
$$
namely pairs of smoothly $\FR^1_t$-parametrized families of a bosonic field configuration and a fermionic field configuration (of linear order)
$$
(\phi_t^{\mathrm{bos.}}\, , \, \psi^\mathrm{fer.}_t\cdot \theta) \quad \in \quad \CF(\FR^{1\vert 1})\, .
$$
\begin{remark}[\bf Super smooth set internal hom]\label{SuperSmoothSetInternalHom}
Precisely the same abstract internal hom mapping space construction as in Rem. \ref{InternalHomMappingSpaceConstruction}, applied now within the category of super smooth sets, recovers exactly the intuitively defined super smooth field spaces of Ex. \ref{Boson-FermionFieldSpacesAsSuperSmoothSets}. See \cite[\S 2.1.5]{GSS24-SuGra} for a discussion towards the example of the gravitino fermion 1-form field of 11D supergravity and \cite{GSc} for further details. 
\end{remark}

\begin{example}[{\bf Jet bundles of odd bundles}]\label{JetBundlesOfOddBundles}
Finite order jet bundles of odd bundles $V_\odd \rightarrow M$ and generally composite bundles $F^{V_\odd}\rightarrow F\rightarrow M$ can be defined (algebraically) as finite dimensional supermanifolds (see \cite{GianchettaGiovanniSardanashvily09}, following \cite{HenneauxTeitelboim92})
$$
J^n_M F^{V_\odd} \quad \in \quad \SuperManifolds \, .
$$
These may be viewed as super smooth sets in the sense of Ex. \ref{SupermanifoldsAsSuperSmoothSets}
$$
y(J^n_M F^{V_\odd}) \quad \in \quad \SuperSmoothSets \, , 
$$
whose limit then defines the infinite jet bundle directly as a super smooth set analogously to the purely smooth case of Ex. \ref{InfiniteJetBundlesAsSmoothSets} 
$$
  y(J^\infty_M F^{V_\odd})
  \quad :=
  \quad \mathrm{lim}_n^{\SuperSmoothSets} y(J^n_M F^{V_\odd})
  \, .
$$ 
This implies that the algebra of finite order functions on $J^\infty_M F^{V_\odd}$
$$
\CO_{\mathrm{glb}}\big(J^\infty_M F^{V^\odd}\big)
$$
exists as maps of super smooth sets, similar to \eqref{GloballyFiniteOrderFunctions}, and hence further 
the variational bicomplex $\Omega^{\bullet,\bullet}(J^\infty_M F^{V_\odd})$ for fermionic field bundles (cf. \cite{GianchettaGiovanniSardanashvily09}) also appears naturally in super smooth sets \cite{GSc}, just as the purely bosonic one did for smooth sets \cite{GS25}. 
\end{example}

With Examples \ref{Boson-FermionFieldSpacesAsSuperSmoothSets} and \ref{JetBundlesOfOddBundles} at hand, the discussion of variations via smooth paths (of plots) of fields from Sec. \ref{SmoothSetsSec} generalizes to arbitrary smooth paths of $\FR^{k\vert q}$-plots of fields, thus including the description of both bosonic and fermionic fields. In particular the corresponding criticality condition (Def. \ref{CriticalRkPoints}) along with the result from \eqref{SmoothSetEulerLagrangeLocus} follow analogously, resulting in a \textit{super smooth set of on-shell fields}. All other field theoretic notions listed in p. 19-20 (as  carefully studied in the purely bosonic case in \cite{GS25}) also extend to the super smooth setting, as necessitated by the existence of fermions. The full details and technicalities of this extension will appear in \cite{GSc}.

\section{Outlook: Infinitesimal and Higher generalizations }\label{GeneralizationSec}
In this closing section, we briefly indicate further generalizations suggested by field theoretic considerations, which are naturally accommodated by further enlarging our collection of probe spaces. 

\subsection*{Infinitesimal structure}

In the theoretical physics literature, one often considers expansions of expressions in 
formal variables $\epsi$ which are taken to be nilpotent in that $\epsi^r=0$ for some $r\in \mathbbm{N}$. For instance, this is the case when defining tangent vectors on manifolds $M$ via ``\textit{infinitesimal curves}  $\gamma_\epsi =\gamma(0)+ \epsi \cdot  \dot{\gamma}(0)$'', or similarly when deriving \textit{infinitesimal} gauge transformations $\delta_\epsi \phi$ from \textit{finite} gauge transformations $\phi\xrightarrow{\, \, g \, \, } \phi'$. A further example is the definition of (on-shell) ``\textit{Jacobi fields}'' as fields that satisfy the linearized field equations, i.e., obtained by formally expanding the full Euler--Lagrange operator and truncating to first order. Of course, another appearance of such manipulations is in (classical) perturbation theory around a fixed field configuration $\phi_0$. Such considerations are rigorously formalized in the context advocated in Sec. \ref{Introduction}, i.e., by probing the \textit{infinitesimal structure} of the field theoretic spaces in question.

Technically, this means we are to further enlarge our collection of probes to include simple infinitesimal spaces. To do this, recall the discussion of p. 24-25, where by ``Milnor's exercise'' we may view our smooth Cartesian probes equivalently as their function algebras  
\begin{align*}
\CartesianSpaces  \quad &\hookrightarrow \quad \mathrm{CAlg}^{\mathrm{op}}_\FR \\
\FR^k \quad &\mapsto \quad C^\infty(\FR^k)\, .
\end{align*}
Now, even prior to passing to super-algebras, we may consider certain commutative algebras as function algebras on would-be infinitesimal spaces. Namely, these are algebras generated by nilpotent even ($\equiv$ bosonic) elements, or more formally quotients of polynomial algebras of the form\footnote{For technical reasons, one needs slightly more general (finite dimensional) nilpotent algebras, known as Weil algebras. These may be equivalently defined as further quotients of the algebras $\CO(\DD^m_l)$.}
$$
\FR[\epsilon_1, ..., \epsilon_m]
    /
    (\epsilon^{r+1}) \quad = : \quad \CO(\mathbbm{D}^m_r)\, .
$$
Such an algebra rigorously encodes the intuition of variables vanishing at some $(r+1)$-polynomial order and so, as the notation on the right suggests, we it take to be the function algebra of the (formal dual) $m$-dimensional \textit{infinitesimal disk (or point)} of order $r$ 
$$
\mathbbm{D}^m_r \quad \in \quad \mathrm{CAlg}^{\mathrm{op}}_\FR \, .
$$

These are the infinitesimal spaces by which we may probe ($\equiv$ define) the infinitesimal structure of our field theoretic spaces. Of course, since we also want to retain the description of smooth structure, we consider our totality of probes to be products of infinitesimal disks with Cartesian spaces, termed \textit{infinitesimally thickened Cartesian spaces}. That is, we take our category category of probes to be  
\begin{align}\label{ThickenedCartesianSpaces}
\mathrm{Th}\CartesianSpaces \quad &\hookrightarrow \quad \mathrm{CAlg}^{\mathrm{op}}_\FR \\
\FR^k\times \DD^m_r \quad &\mapsto \quad \CO(\FR^k \times \DD^m_r) := C^\infty(\FR^k)\otimes \CO(\DD^m_r)\, , \nonumber
\end{align}
which we supply with the (differentiably) good open coverage extended trivially along the infinitesimal directions, analogously to Eq. \eqref{SuperGoodOpenCoverage}. Finally, we define the category of \textit{infinitesimally thickened smooth sets} as the corresponding sheaf category
$$
\mathrm{ThSmoothSets} \quad := \quad \mathrm{Sh}(\mathrm{Th}\CartesianSpaces) \, .
$$
This sheaf topos\footnote{ This is originally due to \cite{Dubuc79}, albeit via a more complicated but equivalent definition of the site of probes (the equivalence being proven in \cite{GSb}). In the mathematics literature, it is usually referred to as the ``\textit{Cahiers topos}''.} forms a \textit{well-adapted model} for \textit{synthetic differential geometry} \cite{Dubuc79}\cite{Kock06}
\noindent \cite{KS17}\cite{GSb}, which is to say that it serves to make the intuitive infinitesimal arguments from physics into rigorous statements.

Let us indicate briefly how this works. Firstly, recall that by the Yoneda embedding \eqref{YonedaEmbedding} manifolds (cf. Ex. \ref{SupermanifoldsAsSuperSmoothSets}) and infinitesimal spaces are seen to inhabit this common category of thickened smooth sets. Paralleling the descriptions of Ex. \ref{FieldSpacesAsSmoothSets} and Ex. \ref{Boson-FermionFieldSpacesAsSuperSmoothSets}, or by using the internal hom of $\mathrm{ThSmoothSets}$ (Rem. \ref{InternalHomMappingSpaceConstruction}),  (bosonic) field spaces $\CF$ may also be seen as such \cite{GSb}
$$
\CF \quad \in \quad \mathrm{ThSmoothSets} \, ,
$$
and similarly for the infinite jet bundles of Ex. \ref{InfiniteJetBundlesAsSmoothSets}. This allows one to probe such generalized spaces by infinitesimal spaces, and in particular the all important (first order) \textit{infinitesimal line}
$$
\DD^1_1 \quad \in \quad \mathrm{Th}\CartesianSpaces \hookrightarrow \mathrm{ThSmoothSets}\, .
$$
\begin{example}[\bf Tangent vectors of manifold are infinitesimal curves]
Consider a smooth manifold $M$. The set of mappings of thickened smooth sets from the infinitesimal line $\DD^1_1$ into $M$ is in canonical bijection with tangent vectors on $M$
$$
TM \quad \cong \quad \mathrm{Hom}_{\mathrm{Th}\SmoothSets}\big( y(\DD^1_1)\, , \, y(M) \big) \,.
$$
Indeed, by the (restricted) Yoneda embedding \eqref{YonedaEmbedding}, such maps of thickened smooth sets are in bijection with algebra homomorphisms $\mathrm{Hom}_{\mathrm{CAlg}_{\FR}}\big( C^\infty(M) \, , \, \CO(\DD^1_1) \big) $.  Any such algebra homomorphism $C^{\infty}(M)\rightarrow \FR[\epsi]/\epsi^2$ 
has components
$$
(p,X_p) \;:\; f \; \longmapsto \; p(f) + \epsi \cdot X_p(f)\, ,
$$
where by $\epsi^2=0$ it follows for $f_1, f_2 \in C^\infty(M)$ that
$$
(f_1 \cdot f_2) \; \longmapsto \; p(f_1)\cdot p(f_2) + \epsi  
\big( p(f_1)\cdot X_p (f_2) + X_p (f_1) \cdot p(f_2) \big)\, .
$$
It follows that the first component $p$ defines the evaluation at a point $p\in M$ (by Milnor's exercise), while the second component $X_p$ defines a derivation at $p\in M$, i.e., a tangent vector.
\end{example}

Using the internal hom construction (Rem. \ref{InternalHomMappingSpaceConstruction}), one can then further explicitly prove (see e.g. \cite{Kock06}) that the \textit{space} of mappings from $\DD^1_1$ into a manifold $M$ recovers the tangent bundle's (thickened) smooth structure 
$$
[y(\DD^1_1)\, , \, y(M)] \quad \cong \quad  y(TM) \quad \in \quad \mathrm{Th}\SmoothSets \, ,
$$
but also that of field spaces \cite{GSb} from Eq. \eqref{FieldSpaceTangentBundleSet} 
$$
[\DD^1_1\, , \, \CF ] \quad \cong \quad T \CF \quad \in \quad \mathrm{ThSmoothSets} \, .
$$
Moreover, if the tangent bundle to the on-shell field space $\CF_{\CE \CL}$ from \eqref{FieldSpaceTangentBundleSet} is defined by the same mapping construction
$$
T\CF_{\CE \CL} \quad := \quad  [\DD^1_1 \, , \, \CF_{\CE \CL}] \, ,
$$
one can prove \cite{GSb} that this recovers also the traditional notion of the (space of) Jacobi fields (as defined for instance in \cite{GS25}). At the level of $*$-plots these identifications can be checked to correspond \textit{precisely} to the naive notion of expanding along an infinitesimal curve, thus fully justifying such intuitive arguments.  

Furthermore, the setting of thickened smooth sets allows for a proper \textit{definition} of an ``\textit{infinitesimal neighborhood}'' $\DD_\phi$ around a point $\phi \in\CF(*)$ of any such space $\CF$. Namely, this is the thickened smooth subset
$$
\DD_\phi \quad \hookrightarrow \quad \CF
$$
whose $(\FR^k\times \DD^m_r)$-plots are given by the subset of $\CF(\FR^k\times \DD^m_r)$ which are constant onto $\phi$ when restricted along $\iota: \FR^k\hookrightarrow \FR^k \times \DD^m_r$. It can be shown explicitly \cite{GSb} that for a manifold $M$ and a point $p\in M$, the restriction of a smooth map
$
S \, : \, M\rightarrow \FR
$
to the infinitesimal neighborhood around $p$
$$
S\vert_{\DD_p} \, : \, \DD_p\hookrightarrow M \longrightarrow \FR
$$
encodes precisely the notion of the formal Taylor expansion, i.e., the perturbative expansion of $S$ around $p$ (w.r.t. any local chart and along with the appropriate notion of the equivalence between expansions corresponding to different charts). We expect that the analogous statement holds for the case of the (infinite dimensional) thickened smooth space of fields $\CF$ corresponding to sections of a field bundle. That is, restriction to the infinitesimal neighborhood around a field configuration $\DD_\phi\hookrightarrow \CF$ should encode the intuitive notion of perturbing around the given field $\phi$ by formally expanding the corresponding action functional / local Lagrangian / Euler--Lagrange operator.

Full details of these and further constructions native to the topos $\mathrm{ThSmoothSets}$ relevant to Lagrangian field theory will appear in \cite{GSb}. Finally, note that including fermionic fields in such constructions necessitates, of course, the introduction of fermionic probes as per Sec. \ref{SuperSmoothSetsSec}. This leads to the category of (thickened) super smooth sets \cite{GSc} as sheaves over (thickened) super Cartesian spaces
\begin{align}\label{ThickenedSuperCartesianSpaces}
\mathrm{Th}\SuperCartesianSpaces \quad &\hookrightarrow \quad \mathrm{SupCAlg}^{\mathrm{op}}_\FR \\
\FR^{k\vert q}\times \DD^m_r \quad &\mapsto \quad \CO(\FR^{k\vert q} \times \DD^m_r) := C^\infty(\FR^k)\otimes\CO(\FR^{0\vert q})\otimes \CO(\DD^m_r)\, . \nonumber
\end{align}

\subsection*{Higher gauge structure} 

Finally, let us return to the familiar \textit{gauge} bosons of the physical world. Recall the standard fact (deduced for instance by considering observables), that the notion of ``sameness'' between any two gauge field configurations $\phi$ and $\phi'$ is not that of equality, but rather being related by a (finite) gauge transformation
$$
\phi \xrightarrow{ \quad g \quad } \phi'
$$
which is in particular invertible. But such a collection of objects (gauge fields) with invertible morphisms between them (finite gauge transformations) is nothing but the definition of a \textit{groupoid} \cite{Weinstein96}. In our probe-wise approach, this translates to the statement that $*$-plots of field configurations of gauge theories should not form mere sets of points, but rather groupoids which naturally encode the \textit{internal} gauge structure of the fields. Similarly then, describing their smooth structure via $\FR^k$-plots should not result into sets, but groupoids of $\FR^k$-parametrized gauge field configurations together with the corresponding gauge transformations.

Before bringing in the smooth structure, note that we can inductively consider ever higher gauge fields (such as the B-field or C-field from supergravity), where gauge transformations themselves are not to be identified by equality but by ``gauge-of-gauge transformations''
\begin{equation}\label{GaugeOfGaugeDiagram}
  \begin{tikzcd}[row sep=small]
    \phi
    \ar[
      rr,
      bend left=30,
      "{ g }",
      "{\ }"{swap, name=s}
    ]
    \ar[
      rr,
      bend right=30,
      "{ g' }"{swap},
      "{\ }"{name=t}
    ]
    \ar[
      from=s,
      to=t,
      Rightarrow, color=greenii,
      "{
        {\color{greenii} h}
      }"
    ]
    &&
    \phi' \, ,
  \end{tikzcd}
\end{equation}
and any two gauge-of-gauge transformations in turn by a higher order gauge transformation
\vspace{-2mm} 
\begin{equation*}
  \begin{tikzcd}[scale=1.5]
    \phi
    \ar[
      rr,
      bend left=50,
      "{ g }",
      "{\ }"{swap, name=s}
    ]
    \ar[
      rr,
      bend right=50,
      "{ g' }"{swap},
      "{\ }"{name=t}
    ]
    \ar[
      from=s,
      to=t,
      shorten=2pt,
      bend left=50,
      Rightarrow, thick, color=greenii,
      "{\ }"{name=tt, swap}, "{
        {\color{greenii} h}
      }"
    ]
    \ar[
      from=s,
      to=t,
      shorten=2pt,
      bend right=50,
      Rightarrow, thick, color=greenii,
      "{\ }"{name=ss}, "{
        {\color{greenii} h'}
      }"{xshift=-13pt}
    ]
    \ar[
      from=ss,
      to=tt,
      phantom, thick, color=orangeii,
      "{
        \Rightarrow
      }"{scale=1.5}
    ]
    &&
    \phi' \, ,
  \end{tikzcd}
\end{equation*}
and so on \textit{ad infinitum}. This necessitates to consider $*$-plots as being instead objects termed ``\textit{$\infty$-groupoids}''. A hands-on approach to $\infty$-groupoids is via the the category of \textit{abstract n-simplices} $\Delta^n$, equipped with the face preserving and collapsing maps \cite[\S 2]{Friedman12}, which encode solely the combinatorial structure of the corresponding geometric simplices (points, lines, triangles, tetrahedra etc.)
$$
\Delta^n_{\mathrm{geo}}
  \;:=\;
  \Big\{
    (x^0, x^1, \cdots, x^n)
    \in 
    \big(
      \mathbb{R}_{\geq 0}
    \big)^n
    \;\Big\vert\;
    \textstyle{\sum_{i = 0}^n} 
    \,
    x^i \,=\, 1
  \Big\} \, .
$$
Viewing the abstract $n$-simplices as probes as per Sec. \ref{Introduction}, one then defines $\infty$-groupoids as certain functors 
$$
\CG \, : \, \Delta^{\mathrm{op}} \xrightarrow{\quad \quad } \mathrm{Set} \, .
$$
Following the terminology of smooth and super sets (Def. \ref{SmoothSetsDefinition} and Def. \ref{SuperSets}), such (pre)sheaves\footnote{That is, sheaves with respect to the trivial coverage on $\Delta$.} are called ``\textit{simplicial sets}''. More precisely, $\infty$-groupoids are defined as those simplicial sets that satisfy a certain existence of (non-unique) higher composites property\footnote{This property is naturally satisfied for field theoretic examples, so we will not worry about it in this exposition.}, called the ``\textit{Kan condition}'' \cite[\S 7]{Friedman12}. Thus $\infty$-groupoids are also known as \textit{Kan simpicial sets} or \textit{Kan complexes}
\begin{equation}\label{InftyGrpdDef}
\mathrm{Grpd}_\infty \, := \, \mathrm{sSet}_{\mathrm{Kan}} \quad \hookrightarrow \quad \mathrm{PreSh}(\Delta) \, .
\end{equation}

The probe-wise description for a given $\infty\mbox{-}\mathrm{groupoid}$ $\CG$ representing the space of higher gauge fields on a spacetime follows as intuitively expected. The $\Delta^0$-plots of $\CG$, i.e., its $0$-simplices is the set
$$
\CG(\Delta^0) \quad = \quad \big\{\, \text{higher gauge field configurations} \,\phi \, \big\} 
$$
of \textit{all} gauge field configurations on a spacetime $M$, possibly including \textit{every  topological sector}, while the set of $1$-simplices
$$
\CG(\Delta^1) \quad = \quad \big\{\, \text{gauge transformations} \, \phi\xrightarrow{\quad g\quad } \phi' \,  \big\} 
$$
is the set of gauge transformations between any two gauge field configurations.
Similarly, the 2-simplices is the set
$$
\CG(\Delta^2) \quad = \quad \big\{\, 2^{nd}\text{-order gauge transformations} \, (g_1,g_2)\xrightarrow{\quad h \quad} g' \,  \big\} 
$$
of (higher) gauge transformations between any pair of consecutive $1^{st}$-order gauge transformations
$$
\phi_1 \xrightarrow{\quad g_1 \quad } \phi_2 \xrightarrow{\quad g_2 \quad } \phi_3 \,
$$
and a third gauge transformation
$$
\phi_1 \xrightarrow{\quad g' \quad } \phi_3 \, , 
$$
thus exhibiting the latter as one possible (gauge equivalent) composition of $g_1$ and $g_2$. A degenerate subcase of these is if $g_2=\id_{\phi_2}$, so that the higher gauge transformation is interpreted between two fixed gauge transformations as depicted in \eqref{GaugeOfGaugeDiagram}.
The pattern continues for $n^{th}$-order higher gauge transformations encoded via the set of $n$-simplices $\CG(\Delta^n)$, for any $n\in \mathbbm{N}$.
\begin{example}[{\bf $\infty$-groupoid of Yang--Mills gauge fields}]\label{InftyGroupoidOfYangMillsGaugeFields}
Consider the classical case of gauge fields of Yang--Mills type for a gauge group $G$ on a spacetime $M$. 
The $0$-simplices of the corresponding $\infty$-groupoid are given by the set of all \textit{global} $G$-gauge field configurations\footnote{Recall its global topological nature is that of a \textit{globally defined} $\mathfrak{g}$-valued 1-form (connection) on a principle $G$-bundle P over $M$, or equivalently via its local transition functions used to glue/identify its \textit{locally defined} $1$-form representatives $\{A_i \in \Omega_\mathrm{dR}^1(U_i, \, \mathfrak{g})\}_{i\in I}$ of $\mathfrak{g}$-valued 1-forms on a cover $\{U_i\hookrightarrow M\}_{i\in I}$ of local charts of $M$ .}
$$
\CG_{M}^{G\mbox{-}\mathrm{gauge}}(\Delta^0) \quad = \quad \big\{\, \text{global G-gauge field configurations}\, A\, \big\}
$$
while the $1$-simplices are given by gauge transformations generated by gauge parameters $g:M\rightarrow \mathrm{Ad}(P)$ into the corresponding adjoint bundles\footnote{These are the bundles $\mathrm{Ad}(P) := P\times_G G$ associated to the adjoint action of the group $G$ on itself. Such globally defined gauge transformations are represented locally by families of $G$-valued functions $\{g_i : U_i \rightarrow G\}$, glued by the same transition functions.}
$$
\CG_{M}^{G\mbox{-}\mathrm{gauge}}(\Delta^1) \quad = \quad \big\{\, \text{gauge transformations}\, \,   A\xrightarrow{\quad g\quad}\, A^g\, :=\,  g^{-1} \cdot A \cdot g + g\cdot  \dd g^{-1} \, \big\} \, .
$$
Since no (non-trivial) higher gauge transformations exist and compositions of gauge transformations are unique (being actually a 1-groupoid), its 2-simplices are tautologically defined and labelled by unique higher (``identity'') morphisms between pairs of gauge transformations and their compositions
$$
\CG_{M}^{G\mbox{-}\mathrm{gauge}}(\Delta^2) \quad = \quad \big\{\,\text{trivial}\, \,   2^{nd}\text{-order gauge transformations} \,  \, (g_1,g_2)\xrightarrow{\quad \mathrm{id}_{g_1\circ g_2}\quad } g_2\circ g_1 \,  \big\} \, .
$$
All higher $n^{th}$-order gauge transformations are also trivial, so the higher simplices $\CG_{M}^{G\mbox{-}\mathrm{gauge}}(\Delta^n)$ follow analogously.
\end{example}
\begin{example}[\bf $\infty$-groupoid of global 2-form gauge fields]\label{InftyGroupoidOfGlobal2FormGaugeFields}
Consider the case of the trivial topological sector of a $B$-field, namely with higher gauge field configurations given by \textit{globally} defined $2$-forms on a spacetime $M$. The $0$-simplices of its $\infty$-groupoid is given by the set of two-forms $\Omega^2_\mathrm{dR}(M)$ 
$$
\CG^{2\mbox{-}\mathrm{form}}_M(\Delta^0) \quad = \quad \big\{\, \text{globally defined 2-forms}\, B\, \big\}
$$
while the $1$-simplices are the gauge transformations generated (and labelled) by globally defined one-forms $A\in \Omega^1_{\mathrm{dR}}(M)$
$$
\CG^{2\mbox{-}\mathrm{form}}_M(\Delta^1) \quad = \quad \big\{\, \text{gauge transformations}\, \,   B\xrightarrow{\quad A\quad}\, B^A\, :=\,  B+ \dd A \, \big\} \, .
$$
Of course, non-trivial higher gauge transformations \textit{do exist} in this case. For instance $A$ and $A+\dd f$ both generate a gauge transformation that maps to the same field configuration
$$
B^{A+\dd f}\, =\,  B+ \dd(A+\dd f) = B+ \dd A\, = \, B^A \, ,
$$
for any $0$-form (function) $f\in \Omega^0_\mathrm{dR}(M)$. More generally, the $2$-simplices are now generated and labelled by $0$-forms $f\in \Omega_{\mathrm{dR}}^0(M)$
$$
\CG^{2\mbox{-}\mathrm{form}}_M(\Delta^2) \quad = \quad \big\{\, 2^{nd}\text{-order gauge transformations} 
 \,\, (A_1,A_2)\xrightarrow{\quad f \quad} A'=A_1+A_2 + \dd f \,  \big\} \, , 
$$
encoding the fact that two consecutive gauge transformations
$$
B_1 \xrightarrow{\quad A_1 \quad } B_2 \, = \, B_1 + \dd A_1 \xrightarrow{\quad A_2 \quad } B_3 \, = \, B_1+ \dd(A_1 + A_2) \,
$$
admit any of the (gauge equivalent) gauge transformations 
$$
B_1 \xrightarrow{\quad A_1+A_2 + \dd f \quad } B_1 + \dd (A_1 + A_2 + \dd f) \, = \, B_3  
$$
as a composed transformation, for any $0$-form $f\in \Omega^0_\mathrm{dR}(M)$. Since no (non-trivial) higher order gauge transformations exist, all the higher simplices $\CG^{2\mbox{-}\mathrm{form}}_M(\Delta^n)$ are tautologically defined and labelled by the corresponding unique (identity) higher morphisms. 
\end{example}

{\bf Smooth and super $\infty$-groupoids}

We now consider reintroducing the smooth structure of Sec. \ref{SmoothSetsSec}, but \textit{along} with the higher gauge structure described above. To start with, this requires that probing the field space with a smooth probe $\FR^k$ yields plots that do not form a set, but rather an $\infty$-groupoid \eqref{InftyGrpdDef}
$$
\CG_{\mathrm{Smth}}(\FR^k) \quad \in \quad \mathrm{sSet}_{\mathrm{Kan}} \, .
$$
for each $\FR^k\in \CartesianSpaces$. In other words, the smooth gauge field space should be defined by a particular kind of functor \cite{dcct}
$$
\CG_{\mathrm{Smth}} \, : \, \CartesianSpaces^{\mathrm{op}} \xrightarrow{\quad \quad}\mathrm{sSet}_{\mathrm{Kan}} \, ,
$$
whose $\FR^k$-plots are interpreted as smoothly $\FR^k$-parametrized families of $\infty$-groupoids comprised of gauge fields, gauge transformations, gauge-of-gauge transformations, and so on. For instance, this means that the $(\FR^k\times \Delta^0)$-plots
\begin{align*}
\CG_{\mathrm{Smth}}(\FR^k\times \Delta^0) \quad  &\equiv \quad  \CG_{\mathrm{Smth}}\big(\FR^k\big)( \Delta^0) \quad \\
&= \quad \big\{\, \text{smoothly}\,  \FR^k\text{-parametrized higher gauge fields} \,\, \phi_{\FR^k} \big\} 
\end{align*}
are the smoothly $\FR^k$-parametrized (higher) gauge field configurations,
while the
$(\FR^k\times \Delta^1)$-plots are the smoothly $\FR^k$-parametrized gauge transformations
\begin{align*}
\CG_{\mathrm{Smth}}(\FR^k\times \Delta^1) \quad  &\equiv \quad  \CG_{\mathrm{Smth}}\big(\FR^k\big)( \Delta^1) \quad \\
&= \quad \big\{\, \text{smoothly}\,  \FR^k\text{-parametrized gauge transformations} \, \, \phi_{\FR^k}\xrightarrow{\quad g_{\FR^k}\quad } \phi_{\FR^k}' \,  \big\} 
\end{align*}
between any two smoothly $\FR^k$-parametrized gauge field configurations, and similarly for higher simplices. Following this intuition, it is not hard to write down the smooth plots corresponding to examples \ref{InftyGroupoidOfYangMillsGaugeFields} and \ref{InftyGroupoidOfGlobal2FormGaugeFields}, following the prescription of Ex. \ref{FieldSpacesAsSmoothSets} at each $\Delta^n$-probe level. The resulting ``smooth $\infty$-groupoids'' are nothing but the finite/integrated version of the corresponding BRST complexes, which are defined by encoding the towers of \textit{infinitesimal} (higher) gauge transformations (dually) as $L_\infty$-algebras (see e.g. \cite{JRSW}\cite{ECPLinfty}). 

Another important class of (higher) smooth spaces that exist in this context are \textit{moduli} or \textit{classifying} spaces of (higher) gauge fields \cite{dcct}.
\begin{example}[\bf Classifying space of G-gauge fields]\label{ClassifyingSpaceOfGgaugeFields}
Recall the $\infty$-groupoid construction of Yang-Mills $G$-gauge fields from Ex. \ref{InftyGroupoidOfYangMillsGaugeFields}, now applied not on a fixed spacetime but instead on \textit{each} Cartesian space $\FR^k$. The  trivial topology of a Cartesian space implies that its objects are thus globally defined 1-forms valued in the Lie algebra $\mathfrak{g}$, and similarly its gauge transformations are globally defined $0$-forms valued in $G$. Explicitly, this defines an assignment of $\infty$-groupoids of $\FR^k$-plots 
$$
\mathbf{B}G_\mathrm{conn} \, : \, \mathrm{CartSp}^{\mathrm{op}} \xrightarrow{\quad \quad} \mathrm{sSet}_{\mathrm{Kan}}
$$
with $0$-simplices, for each $\FR^k\in \CartesianSpaces$,
$$
\mathbf{B}G_\mathrm{conn}\big(\FR^k\big)( \Delta^0) \quad = \quad  \Omega^1_\mathrm{dR}(\FR^k \, , \, \mathfrak{g} )
$$
and corresponding $1$-simplices
$$
\mathbf{B}G_\mathrm{conn}\big(\FR^k\big)( \Delta^1) \quad = \quad  \big\{\, \text{gauge transformations}\, \,    A_{\FR^k}\xrightarrow{\quad g\quad}\, A_{\FR^k}^{g}\, :=\,  g^{-1} \cdot A_{\FR^k} \cdot g + g\cdot  \dd g^{-1} \, \big\} \, .
$$
The higher simplices consist of identity morphisms as in Ex. \ref{InftyGroupoidOfYangMillsGaugeFields}. 
\end{example}
Although this higher smooth space is not interpreted as the space of gauge fields on a fixed spacetime $M$, it instead \textit{classifies} or \textit{modulates} $G$-gauge fields on any spacetime $M$. That is, the set of maps -- in an appropriate sense -- from any manifold $M$ into $\mathbf{B}G_\mathrm{conn}$ is in canonical bijection with the set of $G$-gauge fields, including all topological sectors on $M$
$$
\big\{\, M \xrightarrow{\quad \quad} \mathbf{B}G_\mathrm{conn} \,  \big\} \quad \cong \quad 
 \big\{\, \text{global G-gauge field configurations}\, A\, \big\} \quad = \quad \CG_{M}^{G\mbox{-}\mathrm{gauge}}(\Delta^0) \quad
$$

Of course, we do not address here  what the generalized (higher) version of the sheaf (glueing) condition (cf. p. 7) for their $\FR^k$-plots should be in this setting, nor in which sense traditional natural transformations as maps between any two would-be smooth $\infty$-groupoids $\CX$ and $\CY$ should count as identifying two (gauge) equivalent higher smooth spaces. At an intuitive level, for instance, since the $\FR^k$-plots are no longer sets but have internal gauge symmetries, we should \textit{not only} be able to glue plots that agree on overlaps $U_i \cap U_j \hookrightarrow \FR^k$, but further those that are (consistently) related by gauge transformations on overlaps, and analogously for glueing gauge transformations on triple overlaps when they are related by gauge-of-gauge transformations and so on.  Along the same lines, any natural transformation $\CX\rightarrow \CY$ should count as exhibiting a smooth (gauge) equivalence if each induced map of $\infty$-groupoids of $\FR^k$-plots 
$$
\CX(\FR^k) \xrightarrow{\quad \quad } \CY(\FR^k)
$$
is \textit{locally} around any point $x\in \FR^k$ a gauge equivalence ($\equiv$ homotopy equivalence) of $\infty$-groupoids. We will not enter into the technicalities towards achieving this here, but let us simply mention that the  (homotopy) category of smooth $\infty$-groupoids (representing the corresponding $\infty$-topos) may be defined by formally adding inverses to such local homotopy equivalences (via simplicial localization) 
$$
\mathrm{SmthGrpd}_\infty \quad := \quad L^{\mathrm{lhe}}\, \mathrm{Func}\big(\CartesianSpaces^{\mathrm{op}}\, , \, \mathrm{sSet}_{\mathrm{Kan}}\big)  \, .
$$
The interested reader may consult \cite{dcct}\cite{FSS23Char}\cite{Schreiber24} for details on this definition as it relates to the specific case of smooth $\infty$-groupoids at hand, and where extensive pointers to literature of the general homotopy / $\infty$-category theory involved may be found.

The resulting \textit{higher sheaf topos} of smooth $\infty$-groupoids hosts the smooth gauge field theoretic spaces and the appropriate notion of morphisms between them. Crucially,  this topos also hosts \textit{classifying spaces for higher gauge fields} generalizing Ex. \ref{ClassifyingSpaceOfGgaugeFields}, which neatly encode both the topological and smooth structure of the corresponding field spaces via the internal hom construction (Ex. \ref{InternalHomMappingSpaceConstruction}). Let us close by mentioning a highly non-trivial physical problem that may be rigorously studied in this setting; The \textit{flux quantization} of higher gauge fields \cite{SS23FQ}\cite{SS24Flux}. This formalism encodes Dirac's flux quantization of the electromagnetic field strength $F_2$, the proposed $K$-theoretic flux quantization of RR-fluxes $(F_{2k+0/1})_{k\in \mathbbm{Z}}$ of 10D IIA/IIB supergravities, and furthermore the recently proposed flux quantization of the $(G_4, G_7)$-flux (the ``$C$-field'') in 11D supergravity via ``Hypothesis $H$'' \cite[\S 2.5]{Sati13}. The latter states that the classifying space of globally defined $(G_4,G_7)$-flux configurations is the familiar $4$-sphere $S^4$, viewed as a smooth $\infty$-groupoid. In particular, the \textit{global topological} sector of such a field configuration is encoded by a map of manifolds from spacetime $M$ into the $S^4$.

Finally, bringing the associated \textit{fermionic} matter fields in the picture requires to augment the above discussion by considering $\infty$-groupoids of $\FR^{k\vert q}$-plots\footnote{ And further $\infty$-groupoids of $(\FR^{k\vert q}\times \DD^m_r)$-plots in order to include the infinitesimal nature.}
$$
\CG_{\mathrm{SupSmth}} \, : \, \SuperCartesianSpaces^{\mathrm{op}} \xrightarrow{\quad \quad}\mathrm{sSet}_{\mathrm{Kan}} \, .
$$
In fact, this turns out to be necessary for the aforementioned flux quantization of the $(G_4, G_7)$-flux (and also that of the $H_3$-flux on the M5-brane) to be consistent with the remaining on-shell field equations of 11D supergravity -- including in particular the famous corresponding self-duality constraints. This has only recently been worked out in \cite{GSS24-SuGra}\cite{GSS24-FluxOnM5}, by employing a specially tailored superspace formulation of the theories.

Apart from fleshing out further technical details of the above higher topos in a manner approachable from theoretical physicists, our aim in \cite{GSSd} is to develop a rigorous variational calculus for such (flux quantized) spaces of higher gauge fields.

\vspace{1cm} 
\noindent {\bf Acknowledgements.} 
The author thanks the organizers of the Corfu Summer Institute 2024 for the stimulating and hospitable environment, and the opportunity to present this work. The author is grateful to Hisham Sati and Urs Schreiber for the fruitful ongoing collaboration on which this exposition is largely based upon, and for comments on an earlier draft of this text. The author acknowledges the support by {\it Tamkeen} under the 
{\it NYU Abu Dhabi Research Institute grant} {\tt CG008}.

\end{document}